\documentclass[aps,prd,preprintnumbers]{revtex4}
\usepackage{graphicx}                              
\usepackage{amsmath,amsfonts}
\graphicspath{{./pictures/}}                       

\usepackage{dsfont}   
\usepackage{epsfig}
\usepackage{amssymb}
\usepackage{amsfonts}
\usepackage{float}
\usepackage{bbm}
\usepackage{psfrag}
\newcommand{\vs}{\vspace}
\newcommand{\hs}{\hspace}

\newcommand{\bdm}{\begin{displaymath}}
\newcommand{\edm}{\end{displaymath}}
\newcommand{\beq}{\begin{equation}}
\newcommand{\eeq}{\end{equation}}
\newcommand{\bea}{\begin{eqnarray}}
\newcommand{\eea}{\end{eqnarray}}
\newcommand{\bit}{\begin{itemize}}
\newcommand{\eit}{\end{itemize}}
\newcommand{\bc}{\begin{center}}
\newcommand{\ec}{\end{center}}
\newcommand{\re}{\relax{\textrm{I\kern-.18em R}}}
\newcommand{\ID}{\mathds{1}}

\newcommand{\fhs}[1]{\mbox{\hs{#1}}}
\newcommand{\ie}{\textit{i.e.}$\;$}

\newcommand{\Dov}{{\mathcal D}^{(ov)}}
\newcommand{\D}{\Dov}

\newcommand{\SD}{{\mathcal D}^{(ov)}}

\newcommand{\Dw}{{\mathcal D}^{(W)}}
\newcommand{\SDw}{{\mathcal D}^{(W)}}

\newcommand{\sumFL}{\sum\limits_{i=1}^{N_f}}

\newcommand{\fermiMat}{{\mathcal M}}

\newcommand{\eq}[1]{Eq.~(\ref{#1})}
\newcommand{\eqs}[2]{Eq.~(\ref{#1}-\ref{#2})}
\newcommand{\sect}[1]{section~\ref{#1}}

\newcommand{\sects}[2]{sections~\ref{#1} and \ref{#2}}

\newcommand{\fig}[1]{Fig.~\ref{#1}}
\newcommand{\tab}[1]{Tab.~\ref{#1}}

\newcommand{\latVolExp}[2]{$#1^3\times #2$}
\newcommand{\latVol}[2]{\ifnum #1=#2 $#1^4$ \else $#1^3\times #2$\fi}
\newcommand{\lattice}[2]{\ifnum #1=#2 $#1^4$-lattice \else $#1^3\times #2$-lattice\fi}
\newcommand{\latticeX}[3]{\ifnum #1=#2 $#1^4$-lattice#3 \else $#1^3\times #2$-lattice#3\fi}
\newcommand{\lattices}[2]{\ifnum #1=#2 $#1^4$-lattice \else $#1^3\times #2$-lattices\fi}

\newcommand{\intD}[1]{\mbox{D}#1\,}
\newcommand{\derive}[2]{\frac{\mbox{d}#1}{\mbox{d}#2}}
\newcommand{\funcInt}{\int \intD{\Phi} \intD{\psi} \intD{\bar\psi}}
\newcommand{\GEV}[1]{#1\,\mbox{GeV}}
\newcommand{\TEV}[1]{#1\,\mbox{TeV}}

\newcommand{\Tr}{\mbox{Tr}}
\newcommand{\RE}{\mbox{Re}}
\newcommand{\IM}{\mbox{Im}}
\newcommand{\Ref}[1]{Ref.~\cite{#1}}
\newcommand{\Refs}[1]{Refs.~\cite{#1}}

\newcommand{\SUU}{$\mbox{SU(2)}_L\times\mbox{U(1)}_Y\mbox{ }$}
\newcommand{\proz}[1]{#1\,\%}
\newcommand{\arctanh}{\mbox{arctanh}}

\newcommand{\OneBosonLoopContEuc}{{\mathcal J}}
\newcommand{\OneGoldstoneLoopContEuc}{{\mathcal I}}

\newcommand{\Nconf}{N_{Conf}}
\newcommand{\lowBound}{m_{H}^{low}(\Lambda)}
\newcommand{\upBound}{m_{H}^{up}(\Lambda)}

\newcommand{\SUtwoTimesUoneY}{{SU(2)_\mathrm{L}\times U(1)_\mathrm{Y}}}

\newcommand{\includeFigSingleMedium}[4]{
\bc
\begin{figure}[htb]
\centering
\includegraphics[width=0.75\textwidth]{#1}
\caption[#4]{#3}
\label{#2}
\vs{-2mm}
\end{figure}
\ec
\vs{-6mm}
}

\newcommand{\includeFigDouble}[5]{
\bc
\begin{figure}[htb]
\centering
\begin{tabular}{cc}
\includegraphics[width=0.48\textwidth]{#1}
&
\includegraphics[width=0.48\textwidth]{#2}
\\
\hs{4mm}(a) & \hs{8mm}(b)  \\
\end{tabular}
\caption[#5]{#4}
\label{#3}
\vs{-2mm}
\end{figure}
\ec
\vs{-6mm}
}

\newcommand{\includeFigDoubleDoubleHere}[7]{
\bc
\begin{figure}[ht!]
\centering
\begin{tabular}{cc}
\includegraphics[width=0.48\textwidth]{#1}
&
\includegraphics[width=0.48\textwidth]{#2}
\\
\hs{4mm}(a) & \hs{8mm}(b)  \\
\includegraphics[width=0.48\textwidth]{#3}
&
\includegraphics[width=0.48\textwidth]{#4}
\\
\hs{4mm}(c) & \hs{8mm}(d)  \\
\end{tabular}
\caption[#7]{#6}
\label{#5}
\vs{-2mm}
\end{figure}
\ec
\vs{-6mm}
}

\newcommand{\includeFigTrippleDouble}[9]{
\bc
\begin{figure}[htb]
\centering
\begin{tabular}{ccc}
\includegraphics[width=0.32\textwidth]{#1}\hs{-3mm}
&
\includegraphics[width=0.32\textwidth]{#2}\hs{-3mm}
&
\includegraphics[width=0.32\textwidth]{#3}\hs{-3mm}

\\
\hs{4mm}(a) & \hs{8mm}(b) & \hs{8mm}(c) \\
\includegraphics[width=0.32\textwidth]{#4}\hs{-3mm}
&
\includegraphics[width=0.32\textwidth]{#5}\hs{-3mm}
&
\includegraphics[width=0.32\textwidth]{#6}\hs{-3mm}
\\
\hs{4mm}(d) & \hs{8mm}(e)  & \hs{8mm}(f) \\
\end{tabular}
\caption[#9]{#8}
\label{#7}
\vs{-2mm}
\end{figure}
\ec
\vs{-6mm}
}

\newcommand{\includeFigTriple}[6]{
\bc
\begin{figure}[htb]
\centering
\begin{tabular}{ccc}
\includegraphics[width=0.32\textwidth]{#1} \hs{-3mm}
&
\includegraphics[width=0.32\textwidth]{#2} \hs{-3mm}
&
\includegraphics[width=0.32\textwidth]{#3} \hs{-3mm}
\\
\hs{4mm}(a) & \hs{8mm}(b) & \hs{8mm}(c) \\
\end{tabular}
\caption[#6]{#5}
\label{#4}
\vs{-2mm}
\end{figure}
\ec
\vs{-6mm}
}

\newcommand{\includeTab}[5]{
\begin{table}[htb]
\centering
\begin{tabular}{#1}
\hline
#2
\hline
\end{tabular}
\caption[#5]{#4}
\label{#3}
\end{table}
}

\newcommand{\includeTabNoHLines}[5]{
\begin{table}[htb]
\centering
\begin{tabular}{#1}
#2
\end{tabular}
\caption[#5]{#4}
\label{#3}
\end{table}
}

\newcommand{\includeTabHERE}[5]{
\begin{table}[h!]
\centering
\begin{tabular}{#1}
\hline
#2
\hline
\end{tabular}
\caption[#5]{#4}
\label{#3}
\end{table}
}

\begin{document}
\preprint{HU-EP-10/08, DESY 10-022}

\title{Upper Higgs boson mass bounds from a chirally invariant\\ lattice Higgs-Yukawa model} 

\author{P. Gerhold$^{a,b}$, K. Jansen$^b$}
\affiliation{$^a$Humboldt-Universit\"at zu Berlin, Institut f\"ur Physik, 
Newtonstr. 15, D-12489 Berlin, Germany\\
$^b$NIC, DESY,\\
 Platanenallee 6, D-15738 Zeuthen, Germany}

\date{February 23, 2010}

\begin{abstract}
We establish the cutoff-dependent upper Higgs boson mass bound by means of direct lattice computations 
in the framework of a chirally invariant lattice Higgs-Yukawa model emulating the same chiral Yukawa coupling 
structure as in the Higgs-fermion sector of the Standard Model. As expected from the triviality picture
of the Higgs sector, we observe the upper mass bound to decrease with rising cutoff parameter $\Lambda$.
Moreover, the strength of the fermionic contribution to the upper mass bound is explored by comparing 
to the corresponding analysis in the pure $\Phi^4$-theory. Our final results on the cutoff-dependent 
upper Higgs boson mass bound are summarized in \fig{fig:UpperMassBoundFinalResult}.
\end{abstract}


\maketitle

\includeFigDouble{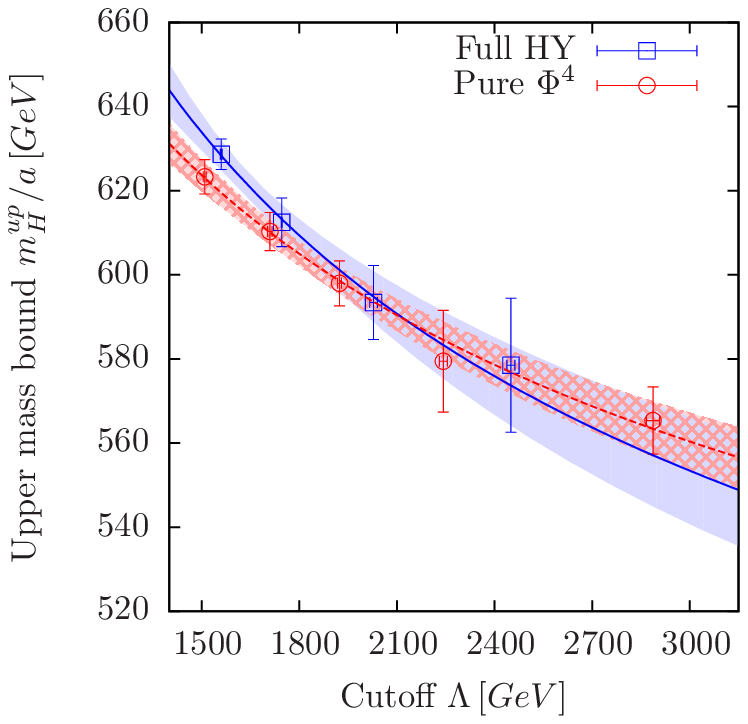}{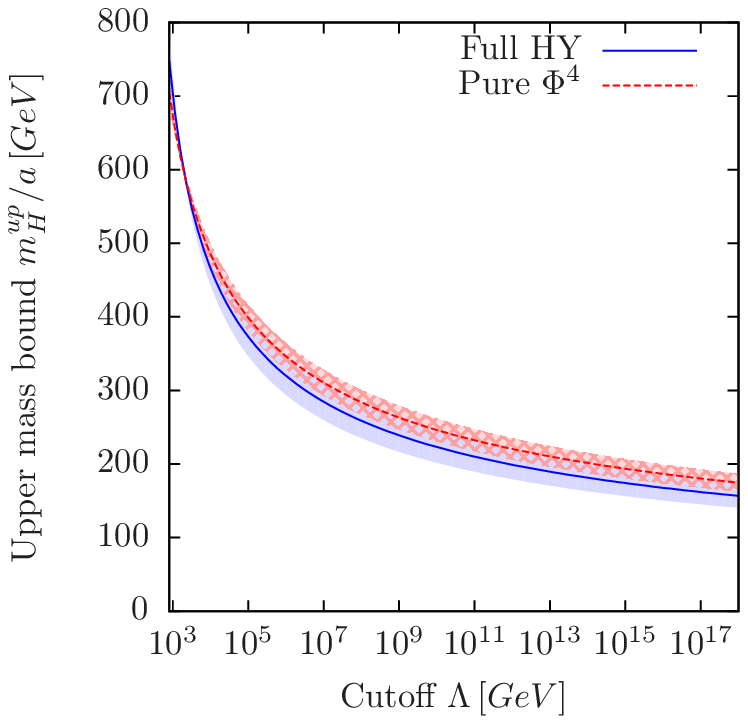}
{fig:UpperMassBoundFinalResult}
{The cutoff dependence of the upper Higgs boson mass bound is presented in panel (a) as obtained from the infinite volume extrapolation results
in \tab{tab:ResultOfUpperHiggsMassFiniteVolExtrapolation1}. The dashed and solid curves are fits of the data arising from the full Higgs-Yukawa model (HY)
and the pure $\Phi^4$-theory, respectively, with the analytically expected cutoff dependence in \eq{eq:StrongCouplingLambdaScalingBeaviourMass}.
Panel (b) shows the aforementioned fit curves extrapolated to larger values of the cutoff $\Lambda$. In both panels the highlighted bands 
reflect the uncertainty of the respective fit curves.
}
{Cutoff dependence of the upper Higgs boson mass bound.}

\section{Introduction}
\label{sec:Introduction}

Given the existing evidence for the triviality of the Higgs sector~\cite{Aizenman:1981du,Frohlich:1982tw,Luscher:1988uq, Hasenfratz:1987eh, Kuti:1987nr, Hasenfratz:1988kr,Gockeler:1992zj} 
of the Standard Model, the latter theory can only be considered as an effective description of Nature valid at most up to some cutoff scale $\Lambda$.
The Higgs sector is thus intrinsically connected with a finite, but unknown cutoff parameter $\Lambda$ that cannot be removed. Beyond that threshold 
an extension of the theory will finally be required. Apriori, the size of this scale $\Lambda$, at which the Standard Model would need such an 
extension, is unspecified. However, the potential discovery of the Higgs boson at the LHC (as well as its non-discovery together with corresponding 
exclusion limits) can shed light on this open question. This can, for instance, be achieved by comparing the experimentally revealed Higgs boson mass 
or its exclusion limits, respectively, with the cutoff-dependent upper and lower Higgs boson mass bounds arising in the Higgs sector of 
the Standard Model.

Besides the obvious interest in narrowing the interval of possible Higgs boson masses consistent with
phenomenology, the latter observation was the main motivation for the great efforts 
spent on the determination of cutoff-dependent upper and lower Higgs boson mass bounds. 
In perturbation theory such bounds have been derived from the criterion of the Landau pole being situated
beyond the cutoff of the theory~(see e.g.~\cite{Cabibbo:1979ay, Dashen:1983ts, Lindner:1985uk}), 
from unitarity requirements~(see e.g.~\cite{Dicus:1992vj, Lee:1977eg, Marciano:1989ns})
and from vacuum stability considerations~(see e.g.~\cite{Cabibbo:1979ay, Linde:1975sw,Weinberg:1976pe,Linde:1977mm, Sher:1988mj, Lindner:1988ww}),
as reviewed in \Ref{Hagiwara:2002fs}.

However, the validity of the perturbatively obtained upper Higgs boson mass bounds is unclear, since
the corresponding perturbative calculations had to be performed at rather large values of the renormalized 
quartic coupling constant. The latter remark thus makes the upper Higgs boson mass bound determination an 
interesting subject for non-perturbative investigations, such as the lattice approach. 

The main objective of lattice studies of the pure Higgs and Higgs-Yukawa sector of the electroweak Standard Model 
has therefore been the non-perturbative determination of the cutoff dependence of the upper Higgs boson mass 
bounds~\cite{Hasenfratz:1987eh, Kuti:1987nr, Hasenfratz:1988kr,Bhanot:1990ai,Holland:2003jr,Holland:2004sd}. 
There are two main developments that warrant the reconsideration of these questions. First, with the advent of the LHC, 
we are to expect that the mass of the Standard Model Higgs boson, if it exists, will be revealed experimentally. 
Second, there is, in contrast to the situation of earlier investigations of lattice 
Higgs-Yukawa  models~\cite{Smit:1989tz,Shigemitsu:1991tc,Golterman:1990nx,book:Jersak,book:Montvay,Golterman:1992ye,Jansen:1994ym}, 
which suffered from their inability to restore chiral symmetry in the
continuum limit while lifting the unwanted fermion doublers at the same
time, a consistent formulation of a Higgs-Yukawa model with an exact 
lattice chiral symmetry~\cite{Luscher:1998pq} based on the Ginsparg-Wilson 
relation~\cite{Ginsparg:1981bj}. This new development allows to maintain the chiral character of the 
Higgs-fermion coupling structure of the Standard Model on the lattice
while simultaneously lifting the fermion doublers, thus eliminating manifestly the main objection to the earlier 
investigations. The interest in lattice Higgs-Yukawa models has therefore recently been
renewed~\cite{Bhattacharya:2006dc,Giedt:2007qg,Poppitz:2007tu,Gerhold:2007yb,Gerhold:2007gx,Fodor:2007fn,Gerhold:2009ub,Gerhold:2010wy}.
In particular, the phase diagram of the new, chirally invariant Higgs-Yukawa model has been discussed analytically by 
means of a large $N_f$ calculation~\cite{Fodor:2007fn, Gerhold:2007yb} as well as numerically by direct Monte-Carlo
computations~\cite{Gerhold:2007gx}. Moreover, the lower Higgs boson mass bounds derived in this lattice model have been 
presented in \Ref{Gerhold:2009ub}. A comprehensive review of these results can be found in \Ref{Gerhold:2010wy}.

In the present paper we intend to determine the dependence of the upper Higgs boson mass bound on the cutoff parameter $\Lambda$ by 
direct Monte-Carlo calculations. In \sects{sec:modelDefinition}{sec:SimStratAndObs} we begin this venture by introducing the 
considered chirally invariant lattice Higgs-Yukawa model and discussing the actual simulation strategy, respectively. Details 
about the determination of the properties of the Goldstone and Higgs boson, in particular their renormalized masses, are then 
given in \sects{sec:GoldPropAnalysisUpperBound}{sec:HiggsPropAnalysisUpperBound}. As a crucial step towards the final 
determination of the upper mass bound we confirm in \sect{sec:DepOfHiggsMassonLargeLam} that the largest Higgs boson masses 
are indeed obtained at infinite bare quartic coupling, as expected from perturbation theory. We then present our results 
on the cutoff dependence of the upper Higgs boson mass bound in \sect{chap:ResOnUpperBound} and examine also the encountered 
finite volume effects. Eventually, the lattice data on the Higgs boson mass bounds are extrapolated to the infinite volume limit, 
yielding then our final result already presented in \fig{fig:UpperMassBoundFinalResult}.

\section{The \SUU lattice Higgs-Yukawa model}
\label{sec:modelDefinition}

The model that will be considered in the following, is a four-dimensional, chirally invariant 
$SU(2)_L \times U(1)_Y$ lattice Higgs-Yukawa model based on the Neuberger overlap operator~\cite{Neuberger:1997fp,Neuberger:1998wv}, 
aiming at the implementation of the chiral Higgs-fermion coupling structure of the pure Higgs-Yukawa 
sector of the Standard Model reading
\beq
\label{eq:StandardModelYuakwaCouplingStructure}
L_Y = y_b \left(\bar t, \bar b \right)_L \varphi b_R 
+y_t \left(\bar t, \bar b \right)_L \tilde\varphi t_R  + c.c.,
\eeq
with $\tilde \varphi = i\tau_2\varphi^*$, $\tau_i$ being the Pauli matrices, and $y_{t,b}$ denoting the bare top and 
bottom Yukawa coupling constants. In this model the consideration is restricted to the top-bottom doublet $(t,b)$ 
interacting with the complex Higgs doublet $\varphi$, which is a reasonable simplification, since the Higgs dynamics 
is dominated by the coupling to the heaviest fermions (apart from its self-coupling). 

The fields contained within the lattice model are thus the scalar field $\varphi$, encoded here however in terms of the 
four-component, real scalar field $\Phi$ for the purpose of a convenient lattice notation, as well as $N_f$ top-bottom 
doublets represented by eight-component spinors $\psi^{(i)}\equiv (t^{(i)}, b^{(i)})$, $i=1,...,N_f$.
In this approach the chiral character of the targeted coupling structure in \eq{eq:StandardModelYuakwaCouplingStructure}
will be preserved on the lattice by constructing the fermionic action $S_F$ from the Neuberger overlap 
operator $\D$ acting on the aforementioned fermion doublets. The overlap operator is given as
\bea
\label{eq:DefOfNeuberDiracOp}
\SD &=& \rho \left\{1+\frac{ A}{\sqrt{ A^\dagger  A}}   \right\},
\quad A = \SDw - \rho, \quad 0 < \rho < 2r
\eea
where $\rho$ is a free, dimensionless parameter within the specified constraints that determines the radius of the circle
formed by the entirety of all eigenvalues of $\D$ in the complex plane. The operator $\SDw$ denotes here the Wilson Dirac 
operator defined as 
\beq
\label{eq:DefOfWilsonOperator}
\Dw = \sum\limits_\mu \gamma_\mu \nabla^s_\mu - \frac{r}{2} \nabla^b_\mu\nabla^f_\mu,
\eeq
where $\nabla^{f,b,s}_\mu$ are the forward, backward and symmetrized lattice nearest neighbor difference operators in 
direction $\mu$, while the so-called Wilson parameter $r$ is chosen here to be $r=1$ as usual.

The overlap operator was proven to be local in a field theoretical sense also in the presence of QCD gauge fields
at least if the latter fields obey certain smoothness conditions~\cite{Hernandez:1998et,Neuberger:1999pz}. The locality properties 
were found to depend on the parameter $\rho$ and the strength of the gauge coupling constant. 
At vanishing gauge coupling the most local operator was shown to be obtained at $\rho=1$. Here, the notion 'most local' 
has to be understood in the sense of the most rapid exponential decrease with the distance $|x-y|$ of the coupling strength 
induced by the matrix elements $\SD_{x,y}$ between the field variables at two remote space-time points $x$ and $y$. For 
that reason the setting $\rho=1$ will be adopted for the rest of this work.

Exploiting the Ginsparg-Wilson relation~\cite{Ginsparg:1981bj} as proposed in \Ref{Luscher:1998pq} one can then 
write down a chirally invariant \SUU lattice Higgs-Yukawa model according to
\bea
\label{eq:DefOfModelPartitionFunctionOriginal}
Z &=& \funcInt e^{-S_\Phi[\Phi] - S_F[\Phi,\psi,\bar\psi]}  \quad \mbox{with} \\
\label{eq:DefYukawaCouplingTerm}
S_F[\Phi,\psi,\bar\psi] &=& \sumFL\,
\bar\psi^{(i)} \underbrace{\left[\D + P_+ \phi^\dagger \fhs{1mm}\mbox{diag}\left(\hat y_t,\hat y_b\right) \hat P_+
+ P_- \fhs{1mm}\mbox{diag}\left(\hat y_t,\hat y_b\right) \phi \hat P_-    \right]}_{\fermiMat}  \psi^{(i)}, \quad\quad\quad
\eea
where the particular form of the $O(4)$-symmetric purely bosonic action $S_\Phi[\Phi]$ will be given later. It is further remarked
that the four-component scalar field $\Phi_x$, defined at the Euclidean site indices $x=(t,\vec x)$ of a $L_s^3\times L_t$-lattice,
has been rewritten here as a quaternionic, $2 \times 2$ matrix $\phi_x = \Phi_x^\mu \theta_\mu,\, \theta_0=\ID,\, \theta_j =  -i\tau_j$  
with $\vec\tau$ denoting the vector of Pauli matrices, acting on the flavour index of the fermion doublets.
The so far unspecified left- and right-handed projection operators $P_\pm$ and their lattice modified counterparts 
$\hat P_{\pm}$ associated to the Neuberger Dirac operator are given as
\bea
P_\pm = \frac{1 \pm \gamma_5}{2},  \quad &
\hat P_\pm = \frac{1 \pm \hat \gamma_5}{2},   \quad &
\hat\gamma_5 = \gamma_5 \left(\ID - \frac{1}{\rho} \D \right).
\eea
The action in \eq{eq:DefYukawaCouplingTerm} now obeys an exact global $\mbox{SU}(2)_L\times \mbox{U}(1)_Y$ 
lattice chiral symmetry. For $\Omega_L\in \mbox{SU}(2)$ and $\epsilon\in \re$ the action is invariant under the transformation
\bea
\label{eq:ChiralSymmetryTrafo1}
\psi\rightarrow  U_R \hat P_+ \psi + U_L\Omega_L \hat P_- \psi,
&\quad&
\bar\psi\rightarrow  \bar\psi P_+ \Omega_L^\dagger U_{L}^\dagger + \bar\psi P_- U^\dagger_{R}, \\
\label{eq:ChiralSymmetryTrafo2}
\phi \rightarrow  U_\phi  \phi \Omega_L^\dagger,
&\quad&
\phi^\dagger \rightarrow \Omega_L \phi^\dagger U_\phi^\dagger
\eea
with $U_{L,R,\phi} \equiv \exp(i\epsilon Y)$ denoting the respective representations of the global $U(1)_Y$ 
symmetry group. Employing the explicit form of the hypercharge $Y$ being related to the isospin component $I_3$ and the 
electric charge $Q$ according to $Y = Q-I_3$, the above $U(1)_Y$ matrices can explicitly be parametrized as 
\bea
U_L = 
\left(
\begin{array}{*{2}{c}}
e^{+i\epsilon/6}&\\
&e^{+i\epsilon/6}\\
\end{array}
\right),
&
U_R = 
\left(
\begin{array}{*{2}{c}}
e^{2i\epsilon/3}&\\
&e^{-i\epsilon/3}\\
\end{array}
\right),
&
U_{\phi} = 
\left(
\begin{array}{*{2}{c}}
e^{+i\epsilon/2}&\\
&e^{-i\epsilon/2}\\
\end{array}
\right),\quad \quad
\eea
for the case of the considered top-bottom doublet. For clarification it is remarked that the right-handed fields are isospin
singlets and have only been written here in form of doublets for the sake of a shorter notation. Note also that in the mass-degenerate 
case, \ie $\hat y_t=\hat y_b$, the above global symmetry is extended to $\mbox{SU}(2)_L\times \mbox{SU}(2)_R$.
In the continuum limit the modified projectors $\hat P_\pm$ converge to $P_\pm$ and the symmetry in \eqs{eq:ChiralSymmetryTrafo1}{eq:ChiralSymmetryTrafo2} 
thus recovers the continuum $\mbox{SU}(2)_L\times \mbox{U}(1)_Y$ global chiral symmetry such that the lattice Higgs-Yukawa coupling becomes 
equivalent to \eq{eq:StandardModelYuakwaCouplingStructure} when identifying 
\bea
\varphi_x = 
C\cdot \left(
\begin{array}{*{1}{c}}
\Phi_x^2 + i\Phi_x^1\\
\Phi_x^0-i\Phi_x^3\\
\end{array}
\right),\quad
&
\tilde\varphi_x = i\tau_2\varphi^*_x = 
C\cdot \left(
\begin{array}{*{1}{c}}
\Phi_x^0 + i\Phi_x^3\\
-\Phi_x^2+i\Phi_x^1\\
\end{array}
\right), \quad
&
y_{t,b} = \frac{\hat y_{t,b}}{C} \quad\quad\quad
\eea
for some real, non-zero constant $C$.
  
The so far unspecified purely bosonic action $S_{\Phi}$ is chosen here to be the lattice version of the $\Phi^4$-action
parametrized in terms of the hopping parameter $\kappa$ and the lattice quartic coupling constant $\hat\lambda$ according to
\beq
\label{eq:LatticePhiAction}
S_\Phi = -\kappa\sum_{x,\mu} \Phi_x^{\dagger} \left[\Phi_{x+\mu} + \Phi_{x-\mu}\right]
+ \sum_{x} \Phi^{\dagger}_x\Phi_x + \hat\lambda \sum_{x} \left(\Phi^{\dagger}_x\Phi_x - N_f \right)^2,
\eeq
which is a convenient parametrization for the actual numerical computations. However, this form of the lattice action 
is fully equivalent to the lattice action in continuum notation 
\bea
\label{eq:BosonicLatticeHiggsActionContNot}
S_{\varphi}[\varphi] &=& \sum\limits_{x,\mu} 
\frac{1}{2} \nabla^f_\mu \varphi^\dagger_x
\nabla^f_\mu \varphi_x
+ \sum\limits_{x} \frac{1}{2} m_0^2 \varphi^\dagger_x\varphi_x
+\sum\limits_{x} \lambda\left(\varphi_x^\dagger \varphi_x \right)^2,
\eea
given in terms of the bare mass $m_0$, the bare quartic coupling constant $\lambda$, and the lattice derivative
operator $\nabla^f_\mu$. The aforementioned connection can be established through a rescaling of the scalar field 
$\Phi$ and the involved coupling constants according to
\beq
\label{eq:RelationBetweenHiggsActions}
\varphi_x = \sqrt{2\kappa}
\left(
\begin{array}{*{1}{c}}
\Phi_x^2 + i\Phi_x^1\\
\Phi_x^0-i\Phi_x^3\\ 
\end{array}
\right), 
\quad
\lambda=\frac{\hat\lambda}{4\kappa^2}, \quad
m_0^2 = \frac{1 - 2N_f\hat\lambda-8\kappa}{\kappa}, \quad
y_{t,b} = \frac{\hat y_{t,b}}{\sqrt{2\kappa}}.
\eeq

Finally, the potential appearance of a sign problem in the framework of the introduced Higgs-Yukawa model
shall be briefly addressed. In the mass-degenerate case, \ie for $y_t=y_b$, one finds that $\det(\fermiMat)\in \re$,
since all eigenvalues of $\fermiMat$ come in complex conjugate pairs according to 
\beq
V\fermiMat V^\dagger = \fermiMat^*, \quad \mbox{with} \quad V = \gamma_0\gamma_2\gamma_5\tau_2.
\eeq
This is in contrast to the general case with $y_t\neq y_b$, where the above relation no longer holds. Throughout 
this work we will therefore only consider the aforementioned mass-degenerate scenario, where the top and bottom
quarks are assumed to have equal masses, to certainly exclude any complex-valued phase of the fermion determinant. 
This, however, still leaves open the possibility of an alternating sign of $\det(\fermiMat)$. We have therefore 
explicitly monitored the sign of $\det(\fermiMat)$ but did never encounter any sign alteration in our actually 
performed Monte-Carlo computations, meaning that the numerical calculations in the mass-degenerate case are 
perfectly sane. A more detailed discussion of the phase of the fermion determinant in the non-degenerate
case can be found in \Ref{Gerhold:2010wy}.

\section{Simulation strategy and considered observables}
\label{sec:SimStratAndObs}

The eventual aim of this work is the non-perturbative determination of the cutoff-dependent upper bound of the Higgs 
boson mass. The general strategy that will be applied for that purpose is to scan through the whole space of bare model 
parameters searching for the largest Higgs boson mass attainable within the pure Higgs-Yukawa sector at a fixed value 
of the cutoff, while being in consistency with phenomenology. This will be done by numerically evaluating the finite lattice model 
of the Higgs-Yukawa sector introduced in the preceding section and extrapolating the obtained results to the infinite 
volume limit. 

The crucial idea is that the aforementioned requirement of reproducing phenomenology restricts the freedom in the choice 
of the bare model parameters $m_0^2, y_{t,b}, \lambda$. For that purpose we exploit here the phenomenological knowledge 
of the renormalized quark masses and the renormalized vacuum expectation value of the scalar field (vev). For the reasons given
in the previous section, however, the top and bottom quarks will be considered to be mass-degenerate. Throughout this work 
$m_{t}/a \equiv m_{b}/a =\GEV{175}$ and $v_r/a = \GEV{246}$ will be assumed. Here $m_t$, $m_b$, and $v_r$ are 
the renormalized top and bottom quark masses as well as the renormalized vev in dimensionless lattice units, while $a$ denotes
the lattice spacing. The aforementioned three conditions leave open an one-dimensional freedom in the bare parameters, which can 
be parametrized in terms of the bare quartic self-coupling constant $\lambda$. However, aiming at the upper Higgs boson mass 
bounds, this remaining freedom can be fixed, since it is expected from perturbation theory that the lightest Higgs boson masses 
are obtained at vanishing self-coupling constant $\lambda=0$, while the heaviest masses are attained at infinite coupling 
constant $\lambda=\infty$, respectively. That this conjecture actually holds also in the non-perturbative regime of the model, 
\ie at large values of $\lambda$, is explicitly demonstrated in \sect{sec:DepOfHiggsMassonLargeLam}, allowing then to restrict 
the search for the upper mass bound to the setting $\lambda=\infty$.

Furthermore, the model has to be evaluated in the broken phase, \ie at $\langle \varphi\rangle \neq 0$, to respect the observation of 
spontaneous symmetry breaking, however close to a second order phase transition to a symmetric phase to allow for arbitrarily large 
correlation lengths as required in any attempt of pushing the cutoff parameter to arbitrarily large values. 

However, in the given lattice model the expectation value $\langle \varphi \rangle$ would always be identical to zero due to the 
symmetries in \eqs{eq:ChiralSymmetryTrafo1}{eq:ChiralSymmetryTrafo2}. The problem is that the lattice averages over {\it all} 
ground states of the theory, not only over that one which Nature has selected in the broken phase. To study 
the mechanism of spontaneous symmetry breaking nevertheless, one usually introduces an external current $J$, selecting then
only one particular ground state. This current is finally removed after taking the thermodynamic limit, leading then to the 
existence of symmetric and broken phases with respect to the order parameter $\langle \varphi\rangle$ as desired.
An alternative approach, which was shown to be equivalent in the thermodynamic 
limit~\cite{Hasenfratz:1989ux,Hasenfratz:1990fu,Gockeler:1991ty}, is to define the vacuum expectation value (vev) $v$ as the 
expectation value of the {\textit{rotated}} field $\varphi^{rot}$ given by a global transformation of the original field $\varphi$ 
according to
\beq
\label{eq:GaugeRotation}
\varphi^{rot}_x = U[\varphi] \varphi_x
\eeq
with the $\mbox{SU}(2)$ matrix $U[\varphi]$ selected for each configuration of field variables $\{\varphi_x\}$
such that 
\beq
\label{eq:GaugeRotationRequirement}
\sum\limits_x \varphi_x^{rot} = 
\left(
\begin{array}{*{1}{c}}
0\\
\left|\sum\limits_x \varphi_x \right|\\
\end{array}
\right).
\eeq
Here we use this second approach. According to the notation in \eq{eq:BosonicLatticeHiggsActionContNot}, which already 
includes a factor $1/2$, the relation between the vev $v$ and the expectation value of $\varphi^{rot}$
is then given as 
\beq
\label{eq:DefOfVEV}
\langle \varphi^{rot} \rangle = 
\left(
\begin{array}{*{1}{c}}
0\\
v\\
\end{array}
\right).
\eeq
In this setup the unrenormalized Higgs mode $h_x$ and the Goldstone modes $g^1_x,g^2_x,g^3_x,$ can then 
directly be read out of the rotated field according to
\beq
\label{eq:DefOfHiggsAndGoldstoneModes}
\varphi_x^{rot}  = 
\left(
\begin{array}{*{1}{c}}
g_x^2 + ig_x^1\\
v + h_x - i g_x^3\\
\end{array}
\right).
\eeq
The great advantage of this approach is that no limit procedure $J\rightarrow 0$ has to be performed, which
simplifies the numerical evaluation of the model tremendously. 

The physical scale of the lattice computation, \ie the inverse lattice spacing $a^{-1}$, can then be determined by comparing the 
renormalized vev $v_r=v/\sqrt{Z_G}$ measured on the lattice with its phenomenologically known value according to
\bea
\label{eq:FixationOfPhysScale}
246\, \mbox{GeV} &=& \frac{v_r}{a} \equiv \frac{v}{\sqrt{Z_G}\cdot a},
\eea
where $Z_G$ denotes the Goldstone renormalization constant. The cutoff parameter $\Lambda$ of the underlying lattice regularization,
which is directly associated to the lattice spacing $a$, can then be defined as
\beq
\label{eq:DefOfCutoffLambda}
\Lambda = a^{-1}.
\eeq

Of course, this definition is not unique and other authors use different definitions, for instance $\Lambda=\pi/a$ motivated 
by the value of the momenta at the edge of the Brillouin zone. However, since the quantities that actually enter any lattice 
calculation are rather the lattice momenta $\tilde p_\mu=sin(p_\mu)$ instead of the momenta $p_\mu$, which are connected through
the application of a sine function, it seems natural to choose the definition of the cutoff $\Lambda$ given in \eq{eq:DefOfCutoffLambda}. 

Next, the extraction technique for the Goldstone renormalization constant entering \eq{eq:FixationOfPhysScale} needs to be 
determined. In the Euclidean continuum the Goldstone and Higgs renormalization constants, more precisely their inverse
values $Z^{-1}_G$ and $Z^{-1}_H$, are usually defined as the real part of the derivative of the inverse Goldstone and Higgs 
propagators in momentum space with respect to the continuous squared momentum $p_c^2$ at some scale $p_c^2=-\mu_G^2$ and 
$p_c^2=-\mu_H^2$, respectively. The restriction to the real part is introduced to make this definition applicable also in 
the case of an unstable Higgs boson, where the massless Goldstone modes induce a branch cut with discontinuous complex contributions
to the propagator at negative values of $p_c^2$. This is the targeted definition that shall also be adopted to the later lattice 
calculations. 

On the lattice, however, the propagators are only defined at the discrete lattice momenta $p_\mu=2\pi n_\mu/L_{s,t}$, $n_\mu = 0, \hdots, L_{s,t}-1$
according to
\bea
\tilde G_H(p) &=& \langle \tilde h_p \tilde h_{-p}\rangle, \\
\tilde G_G(p) &=& \frac{1}{3}\sum\limits_{\alpha=1}^3 \langle \tilde g^\alpha_p \tilde g^\alpha_{-p}\rangle, 
\eea
where the Higgs and Goldstone fields in momentum representation read
\bea
\tilde h_p = \frac{1}{\sqrt{V}}\sum\limits_x e^{-ipx} h_x &\mbox{ and }&
\tilde g^\alpha_p = \frac{1}{\sqrt{V}}\sum\limits_x e^{-ipx} g^\alpha_x
\eea
with $V=L_s^3\cdot L_t$ denoting the lattice volume.

Computing the derivative of the lattice propagators is thus not a well-defined operation. Moreover, the lattice
propagators are not even functions of $p^2$, since rotational invariance is explicitly broken by the discrete
lattice structure. To adopt the above described concept to the lattice nevertheless, some lattice scheme has to be
introduced that converges to the continuum definitions of $Z_G$ and $Z_H$ in the limit $a\rightarrow 0$.
Here, the idea is to use some analytical fit formulas $f_{G}(p)$, $f_{H}(p)$ derived from renormalized 
perturbation theory in the Euclidean continuum to approximate the measured lattice propagators $\tilde G_G(p)$ and 
$\tilde G_H(p)$ at small momenta $\hat p^2<\gamma$ (with $\hat p^2_\mu \equiv 4\sin^2(p_\mu/2)$) for some appropriate 
value of $\gamma$ such that the discretization errors are acceptable. The details of this fit procedure are discussed 
in \sects{sec:GoldPropAnalysisUpperBound}{sec:HiggsPropAnalysisUpperBound}. One can then define the analytically 
continued lattice propagators as
\bea
\tilde G_G^{c}(p_c) = f_{G}(p_c) &\mbox{ and }&
\tilde G_H^{c}(p_c) = f_{H}(p_c).
\eea
In the on-shell scheme the targeted Goldstone and Higgs renormalization constants $Z_G$ and $Z_H$ can then be 
defined (implicitly assuming an appropriate mapping $p_c\leftrightarrow p_c^2$) as 
\bea
Z^{-1}_G(\mu^2_G) &=& \derive{}{p_c^2} \RE\left(\left[\tilde G^{c}_G(p_c^2)\right]^{-1}\right)\Big|_{p_c^2 = -\mu^2_G},\\
Z^{-1}_H(\mu^2_H) &=& \derive{}{p_c^2} \RE\left(\left[\tilde G^{c}_H(p_c^2)\right]^{-1}\right)\Big|_{p_c^2 = -\mu^2_H},
\eea
with $\mu^2_G=m^2_G$ and $\mu^2_H=m^2_H$, where the underlying physical masses $m_G$, $m_H$ are given by the poles of the respective 
propagators on the second Riemann sheet. To adopt this definition to the introduced lattice scheme we define the Higgs 
boson mass $m_H$, its decay width $\Gamma_H$, and the mass $m_G$ of the stable Goldstone bosons through
\bea
\label{eq:DefOfHiggsAndGoldstoneMassByPole}
\left[\tilde G^{c}_{H,II}(im_H+\Gamma_H/2,0,0,0)\right]^{-1} = 0, &\mbox{ and }&
\left[\tilde G^{c}_G(im_G,0,0,0)\right]^{-1} = 0,
\eea
where $\tilde G^{c}_{H,II}(p_c)$ denotes the analytical continuation of $\tilde G^{c}_{H}(p_c)$ onto the second Riemann sheet.

Extracting the Higgs boson mass $m_H$ and its decay width $\Gamma_H$ from simulation data according to this definition would, 
however, require an explicit analytical continuation of the Higgs propagator onto the second Riemann sheet, which is 
beyond our ambitions in this study. 

Following the proposal in \Ref{Luscher:1988uq} the Goldstone and Higgs renormalization factors are rather determined 
at the scales $\mu_{G}^2 = m^2_{Gp}$ and $\mu_{H}^2 = m^2_{Hp}$ given by the masses $m_{Hp}$ and $m_{Gp}$, which 
will be referred to in the following as propagator masses in contrast to the pole masses $m_H$ and $m_G$. 
We thus define
\bea
\label{eq:DefOfRenormalFactors}
Z_G \equiv Z_G(m^2_{Gp})  &\mbox{ and }&
Z_H \equiv Z_H(m^2_{Hp}),
\eea
where the propagator masses $m_{Hp}$, $m_{Gp}$ are defined through a vanishing real-part of the inverse 
propagators according to
\bea
\label{eq:DefOfPropagatorMinkMass}
\RE\left(\left[\tilde G_G^{c}(p_c^2)\right]^{-1}\right)\Big|_{p_c^2 = -m^2_{Gp}} = 0 &\mbox{ and }& 
\RE\left(\left[\tilde G_H^{c}(p_c^2)\right]^{-1}\right)\Big|_{p_c^2 = -m^2_{Hp}} = 0.
\eea
The reasoning for selecting these latter definitions of the Higgs and Goldstone masses is, that the required 
analytical continuation in the case of the Higgs propagator is much more robust, since it only needs to extend 
the measured lattice propagator to purely negative values of $p_c^2$ in contrast to the situation resulting from 
the definition in \eq{eq:DefOfHiggsAndGoldstoneMassByPole}. It is remarked here that the Goldstone propagator
mass $m_{Gp}$ was only introduced for the sake of an uniform notation, since $m_{G}$ is identical to $m_{Gp}$, 
due to the Goldstone bosons being stable particles.

As for the unstable Higgs boson, however, one finds that the discrepancy between the pole mass $m_H$ and the
propagator mass $m_{Hp}$ is directly related to the size of the decay width $\Gamma_H$. In the weak coupling 
regime of the theory the two mass definitions $m_H$ and $m_{Hp}$ can thus be considered to coincide up to
small perturbative corrections, due to a vanishing decay width in that limit. For the pure $\Phi^4$-theory
the deviation between $m_{Hp}$ and $m_H$ has explicitly been worked out in renormalized perturbation theory~\cite{Luscher:1988uq}. 
In infinite volume the finding is
\beq
\label{eq:ConnectionMhMhp}
m_H = m_{Hp} \cdot \left(1+\frac{\pi^2}{288} (n-1)^2 \left[\frac{4!\cdot\lambda_r}{16\pi^2}\right]^2  + O(\lambda_r^3)  \right),
\eeq
where $\lambda_r$ denotes the renormalized quartic self-coupling constant and $n$ is the number of components 
of the scalar field $\Phi$, \ie $n=4$ for the here considered case. This calculation was performed in 
the pure $\Phi^4$-theory, thus neglecting any fermionic degrees of freedom, and for exactly massless 
Goldstone particles. However, one learns from this result that the definition of $m_{Hp}$ in \eq{eq:DefOfPropagatorMinkMass} 
as the Higgs boson mass is very reasonable at least for sufficiently small values of the renormalized coupling constants.
The actual discrepancy between $m_H$ and $m_{Hp}$ as obtained by direct lattice computations of their respective
definitions in \eq{eq:DefOfHiggsAndGoldstoneMassByPole} and \eq{eq:DefOfPropagatorMinkMass} will 
explicitly be examined in \sect{sec:HiggsPropAnalysisUpperBound} for some physically relevant parameter setups.
It will then indeed be found to be negligible with respect to the reachable statistical accuracy. 

The definition of the renormalized quartic self-coupling constant $\lambda_r$ that was used in the derivation of \eq{eq:ConnectionMhMhp}
is
\beq
\label{eq:DefOfRenQuartCoupling}
\lambda_r = \frac{m^2_{Hp}-m^2_{Gp}}{8v_r^2},
\eeq
which shall also be taken over to the considered Higgs-Yukawa model. 
In principle, it would also be possible to determine the renormalized quartic coupling constant $\lambda_r$ through
the evaluation of the amputated, connected, one-particle-irreducible four-point function at a specified momentum configuration 
as it is usually done in perturbation theory. However, the signal to noise ratio of the corresponding lattice observable is suppressed by the
lattice volume. It is thus extremely hard to measure the renormalized quartic coupling constant in lattice calculations by means
of the direct evaluation of such four-point functions~\cite{Jansen:1988cw}. Instead, the alternative definition of $\lambda_r$ 
given in \eq{eq:DefOfRenQuartCoupling} will be adopted here. It is further remarked that this definition was shown~\cite{Luscher:1988uq} 
to coincide with the bare coupling parameter $\lambda$ to lowest order in the pure $\Phi^4$-theory. 

Regarding the top and bottom quark fields, we are here only interested in the corresponding masses $m_t, m_b$. These can directly be obtained 
by studying the fermionic time correlation functions $C_f(\Delta t)$ at large Euclidean time separations $\Delta t$, where 
$f=t,b$ denotes the quark flavour here. On the lattice the latter time correlation functions can be defined as
\bea
\label{eq:DefOfFermionTimeSliceCorr}
C_f(\Delta t) &=& \frac{1}{L_t\cdot L_s^6} \sum\limits_{t=0}^{L_t-1} \sum\limits_{\vec x, \vec y}
\Big\langle 2\,\RE\,\Tr\,\left(f_{L,t+\Delta t, \vec x}\cdot \bar f_{R,t,\vec y}\right) \Big\rangle,
\eea
where the left- and right-handed spinors are given through the projection operators according to
\bea
\label{eq:DefOfLeftHandedSpinors}
\left(
\begin{array}{*{1}{c}}
t\\
b\\
\end{array}
\right)_L
= \hat P_- 
\left(
\begin{array}{*{1}{c}}
t\\
b\\
\end{array}
\right)
 &\mbox{ and }&  (\bar t, \bar b)_R  = (\bar t, \bar b) P_-.
\eea
It is remarked that the given fermionic correlation function would 
be identical to zero due to the exact lattice chiral symmetry obeyed by the considered Higgs-Yukawa model, if one would not
rotate the scalar field $\varphi$ according to \eq{eq:GaugeRotation} as discussed above. This 
rotation is implicitly assumed in the following. Furthermore, it is pointed out that the full {\textit{all-to-all}} correlator 
as defined in \eq{eq:DefOfFermionTimeSliceCorr} can be trivially computed. This all-to-all correlator yields very clean 
signals for the top and bottom quark mass determination.

The lacking definition of the renormalized Yukawa coupling constants can now be provided as
\bea
\label{eq:DefOfRenYukawaConst}
y_{t,r} = \frac{m_t}{v_r} &\mbox{ and }&
y_{b,r} = \frac{m_b}{v_r},
\eea
reproducing the bare Yukawa coupling constants $y_{t,b}$ at lowest order. According to the presented simulation strategy 
the aim would thus be to tune the above renormalized Yukawa coupling constants such that their physically known values would
be reproduced in the actual lattice computations. However, for having some initial guess for the latter adjustment at hand 
the tree-level relation 
\beq
\label{eq:treeLevelTopMass}
y_{t,b} = \frac{m_{t,b}}{v_r}
\eeq
will be used throughout this work to set the bare Yukawa coupling constants in the lattice computations. Comparing the physical 
fermion masses actually generated in these lattice calculations with the targeted ones would then allow to fine tune the Yukawa 
coupling constants in an iterative refinement approach. However, it turns out that this tree-level fixation ansatz already yields 
quite satisfactory results regarding the discrepancy between the targeted and the actually observed quark masses with respect 
to the reached statistical accuracy.

\section{Analysis of the Goldstone propagator}
\label{sec:GoldPropAnalysisUpperBound}

The Goldstone renormalization constant $Z_G$ is required for determining the renormalized vacuum expectation
value $v_r$ of the scalar field. It is thus needed for the fixation of the physical scale $a^{-1}$ of a given Monte-Carlo run
according to \eq{eq:FixationOfPhysScale}. This renormalization constant has been defined in \eq{eq:DefOfRenormalFactors} 
through a derivative of the inverse Goldstone propagator. As already pointed out in \sect{sec:SimStratAndObs} computing 
this derivative requires an analytical continuation $\tilde G_G^{c}(p_c)$ of the discrete lattice propagator, which was proposed to be 
obtained via a fit of the discrete lattice data. 

\includeFigSingleMedium{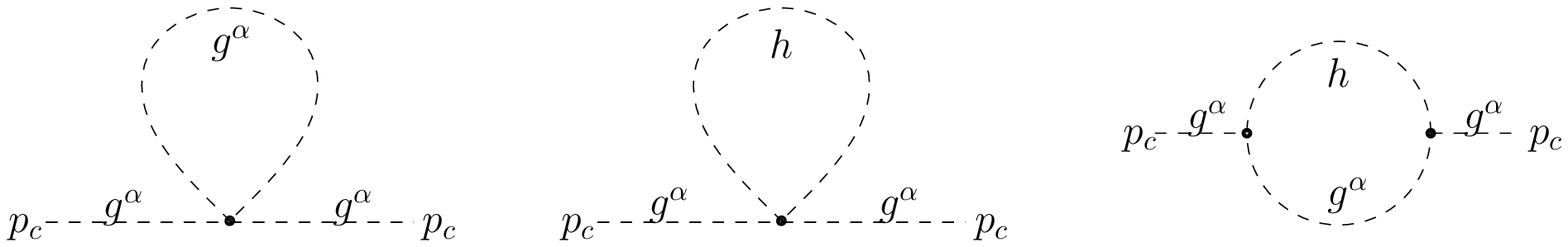}
{fig:GoldstonePropDiagrams}
{Illustration of the diagrams that contribute to the continuous space-time Goldstone propagator $\tilde G_G(p_c)$
in the Euclidean pure $\Phi^4$-theory at one-loop order.
}
{Diagrams contributing to the continuous space-time Goldstone propagator in the pure $\Phi^4$-theory at one-loop order.}

The idea here is to construct an appropriate fit function $f_G(p)$ based on a perturbative calculation of the Goldstone
propagator $\tilde G_G(p_c)$ in continuous Euclidean space-time. In this study the aforementioned fit function will only 
play the role of an effective description of the numerical data to allow for the necessary analytical continuation. For its 
construction we can therefore impose a set of simplifications. In particular, we restrict the consideration here to the pure 
$\Phi^4$-theory. The reasoning behind this simplification is that the purely bosonic four-point interaction is
expected to yield the dominant contributions to the Goldstone propagator in the targeted strong coupling regime with
infinite bare $\lambda$ but only moderate values of the bare Yukawa coupling constants. To one-loop order the only momentum dependent 
contribution to the Goldstone propagator is thus given by the mixed Higgs-Goldstone loop illustrated on the right-hand 
side of \fig{fig:GoldstonePropDiagrams}, where the system has been assumed to be in the broken phase, as desired. At 
one-loop order the result for the renormalized Goldstone propagator then reads
\bea
\label{eq:GoldstonePropRenPT}
\tilde G^{-1}_G(p_c) &=& p_c^2 + m^2_{G} + 8\pi^{-2}\lambda_r^2 v_r^2 \cdot \left[
\OneGoldstoneLoopContEuc(p_c^2,m_H^2,m_G^2)  - \OneGoldstoneLoopContEuc(-m_{G}^2,m_H^2,m_G^2)\right] 
\eea
where the one-loop contribution $\OneGoldstoneLoopContEuc(p_c^2, m_H^2, m_G^2)$ is given as
\bea
2\OneGoldstoneLoopContEuc(p_c^2,m_H^2,m_G^2) &=& \frac{\sqrt{q}}{p_c^2} \cdot \arctanh\left( \frac{p_c^2+m^2_{G}-m^2_{H}}{\sqrt{q}} \right) 
+ \frac{m^2_{G} - m^2_{H}}{2p_c^2}\cdot \log\left( \frac{m^2_{H}}{m^2_{G}}\right)\quad \quad \\
&+&\frac{\sqrt{q}}{p_c^2} \cdot \arctanh\left(\frac{p_c^2+m^2_{H}-m^2_{G}}{\sqrt{q}}\right) \quad \mbox{with}\nonumber\\
q &=& \left(m^2_{G} - m^2_{H} + p_c^2  \right)^2 + 4m^2_{H} p_c^2.
\eea
Concerning the singularities of this expression it is noteworthy to add that the given formula can be shown to be finite at 
$p_c=0$ for $m_G\neq 0$, as desired. 

In principle, one can directly employ the expression in \eq{eq:GoldstonePropRenPT} as the sought-after fit function $f^{-1}_G(p)$. 
For clarification it is remarked at this point that instead of fitting the lattice propagator $\tilde G_G(p)$ itself with $f_G(p)$,
it is always the inverse propagator $\tilde G_G^{-1}(p)$ that is fitted with $f_G^{-1}(p)\equiv 1/f_G(p)$ in the following. However, 
for the actual fit procedure of the lattice data a modified version of \eq{eq:GoldstonePropRenPT} is used given as
\beq
\label{eq:GoldstoneFitAnsatz}
f^{-1}_G(p^2) = \frac{p^2 + \bar m^2_{G} +  A\cdot \left[ \OneGoldstoneLoopContEuc(p^2,\bar m_H^2, \bar m_G^2)  - 
\OneGoldstoneLoopContEuc(0,\bar m_H^2, \bar m_G^2)\right]}{Z_0},
\eeq
where an appropriate mapping $p^2\leftrightarrow p$ is implicit and $A$, $Z_0$, $\bar m_G$, $\bar m_H$ are the free fit parameters. 
Two modifications have been applied here to the original result. 
Firstly, the constant term $\OneGoldstoneLoopContEuc(-\bar m_{G}^2,\bar m_H^2,\bar m_G^2)$ in \eq{eq:GoldstonePropRenPT} has been replaced
by $\OneGoldstoneLoopContEuc(0,\bar m_H^2,\bar m_G^2)$ simply for convenience. Since the Goldstone mass is close to zero anyhow, this 
simplification is insignificant for a practical fit procedure. For clarification it is recalled that in the presented approach the 
Goldstone mass $m_G$ is actually not determined through the nominal value of the latter fit parameter $\bar m_G$ itself, but through 
the pole of the resulting analytical continuation $\tilde G_G^{c}(p_c)$ according to \eq{eq:DefOfHiggsAndGoldstoneMassByPole}. This is
also indicated by the chosen notation introducing the symbol $\bar m_G$ in addition to the actual Goldstone mass $m_G$.

More interestingly, however, a global factor $Z_0$ has been included in the denominator of \eq{eq:GoldstoneFitAnsatz} in the 
spirit of a renormalization constant. This modification is purely heuristic and its sole purpose is to provide an effective 
description of the so-far neglected fermionic contributions, which is all we need at this point. 

Of course, it would be more appropriate to construct a fit ansatz from the renormalized result of the Goldstone
propagator derived in the full Higgs-Yukawa sector. This would indeed place the fit procedure on an even better conceptual footing.
However, it will turn out, that the given ansatz already works satisfactorily well for our purpose, which is not too surprising,
due to the aforementioned dominance of the quartic coupling term in that model parameter space being of physical 
interest here.

More important seems to be the question what part of the lattice Goldstone propagator $\tilde G_G(p)$ one should actually include
into the fit procedure. It was already pointed out in \sect{sec:SimStratAndObs} that the consideration of the lattice propagator 
has to be restricted to small lattice momenta in order to suppress contaminations arising from discretization effects. For that purpose
the constant $\gamma$ has been introduced specifying the set of momenta underlying the fit approach according to $\hat p^2 \le \gamma$.
In principle, one would want to choose $\gamma$ as small as possible. In practice, however, the fit procedure becomes increasingly
unstable when lowering the value of $\gamma$, because less and less data are then included within the fit. In the following example
lattice computations, demonstrating the evaluation approach for $Z_G$ and $m_G$, we will consider the settings $\gamma=1$, 
$\gamma=2$, and $\gamma=4$. To make the discretization effects associated to these not particularly small values of $\gamma$ less 
prominent in the intended fit procedure, the inverse lattice propagator $\tilde G^{-1}_G(p)$ is actually fitted with $f^{-1}_G(\hat p^2)$ 
instead of $f^{-1}_G(p^2)$, being a function of the squared lattice momentum $\hat p^2$, which is completely justified in the 
limit $\gamma\rightarrow 0$.

\includeTab{|cccccccc|}
{
\latVolExp{L_s}{L_t} & $N_f$ & $\kappa$  & $\hat \lambda$ & $\hat y_t$     & $\hat y_b/\hat y_t$ & $v$ & $\Lambda$ \\
\hline
\latVol{32}{32}      & $1$   & $0.30039$ & $\infty$       & $0.55139$ & $1$       & $ 0.1008(3)\,\,$ & $\GEV{ 2373.0\pm 6.4}$\\
\latVol{32}{32}      & $1$   & $0.30400$ & $\infty$       & $0.55038$ & $1$       & $ 0.1547(1)\,\,$ & $\GEV{ 1548.1\pm 1.8}$\\
}
{tab:Chap71EmployedRuns}
{The model parameters of the Monte-Carlo runs constituting the testbed for the subsequently discussed computation schemes 
are presented together with the obtained values of the vacuum expectation value $v$ and the cutoff $\Lambda$
determined by \eq{eq:DefOfCutoffLambda}. The degenerate Yukawa coupling constants have been chosen here according
to the tree-level relation in \eq{eq:treeLevelTopMass} aiming at the reproduction of the phenomenologically known top 
quark mass.
}
{Model parameters of the Monte-Carlo runs constituting the testbed for the considered computation schemes at large quartic
coupling constants.}

The Goldstone propagators obtained in the lattice calculations specified in \tab{tab:Chap71EmployedRuns} are 
presented in \fig{fig:GoldstonePropExampleArStrongCoup}. These numerical data of the inverse Goldstone propagator $\tilde G_G^{-1}(p)$ 
have been fitted with the fit formula $f^{-1}_G(\hat p^2)$ given in \eq{eq:GoldstoneFitAnsatz}.
One can observe already from the graphical presentation in \fig{fig:GoldstonePropExampleArStrongCoup} that the considered
fit ansatz $f_G(\hat p^2)$ describes the numerical data significantly better than the simple linear fit formula 
\beq
\label{eq:GoldstoneFitAnsatzLinear}
l_G^{-1}(\hat p^2) = \frac{\hat p^2 + m_G^2}{Z_G},
\eeq
which is additionally considered here for the only purpose of demonstrating the quality of the applied fit ansatz $f_G(\hat p^2)$.

To find an optimal setting for the threshold value $\gamma$, the dependence of the fit results on the latter parameter is listed
in \tab{tab:GoldstonePropExampleResultsAtStrongCoup}, where the presented Goldstone mass $m_G$ and the renormalization 
factor $Z_G$ have been obtained according to \eq{eq:DefOfHiggsAndGoldstoneMassByPole} and \eq{eq:DefOfRenormalFactors} from the 
analytical continuation of the lattice Goldstone propagator given by $\tilde G^{c}_G(p_c) = f_G(p_c)$ and $\tilde G^{c}_G(p_c) = l_G(p_c)$, 
respectively. 
 
At first glance one notices that the linear ansatz $l_G(\hat p^2)$ yields more stable results than $f_G(\hat p^2)$. These
results are, however, inconsistent with themselves when varying the parameter $\gamma$. One can also observe
in \tab{tab:GoldstonePropExampleResultsAtStrongCoup} that the associated average squared residual per degree of freedom 
$\chi^2/dof$ significantly differs from one at the selected values of $\gamma$, making apparent that the simple linear
fit ansatz is not suited for the reliable determination of the Goldstone propagator properties.

\includeFigTrippleDouble
{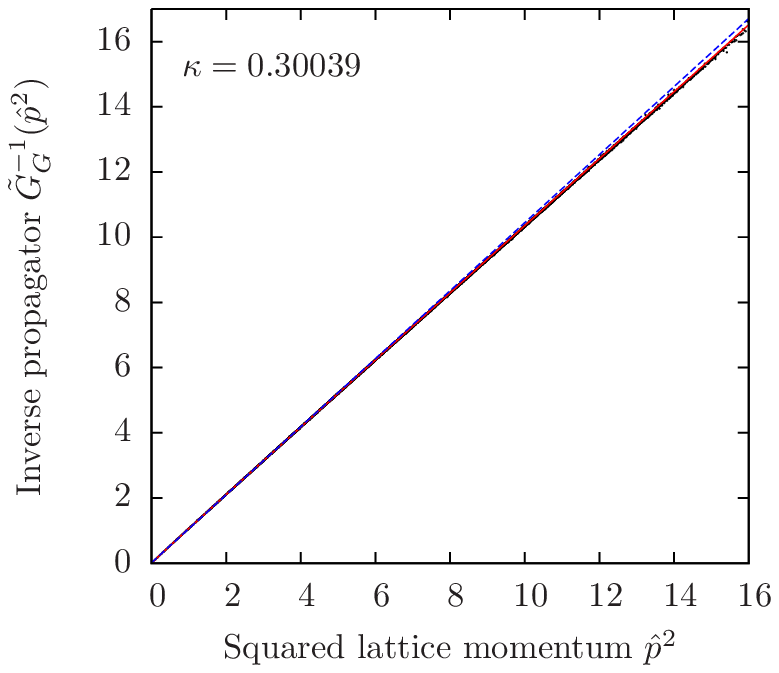}{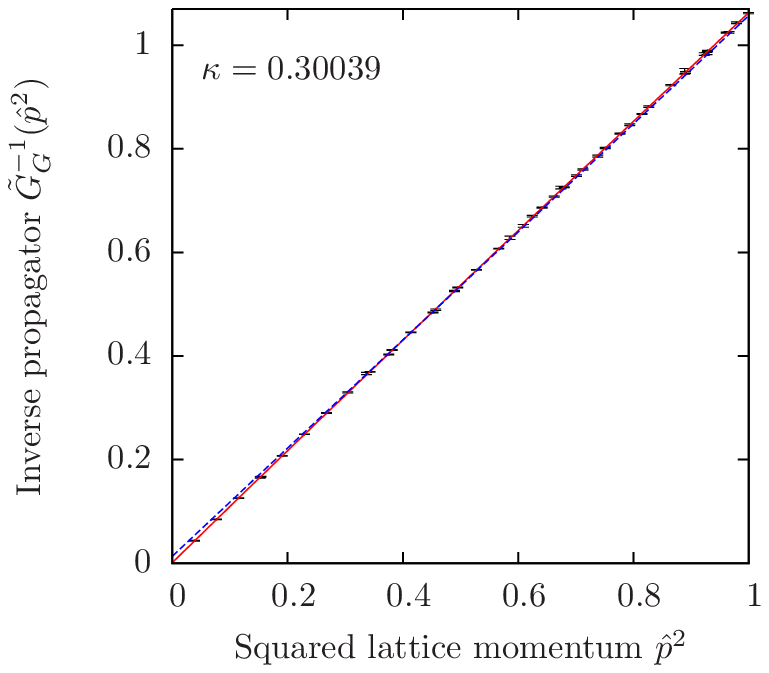}{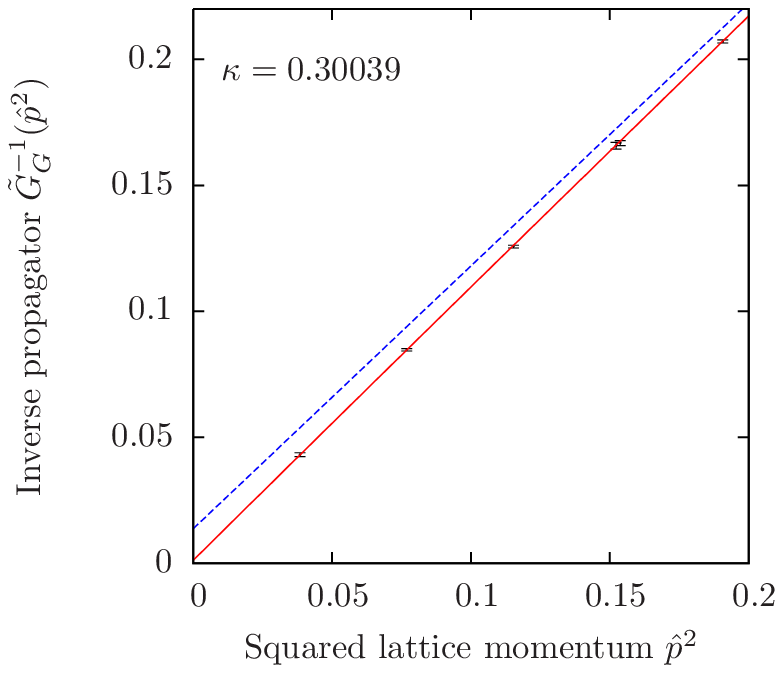}
{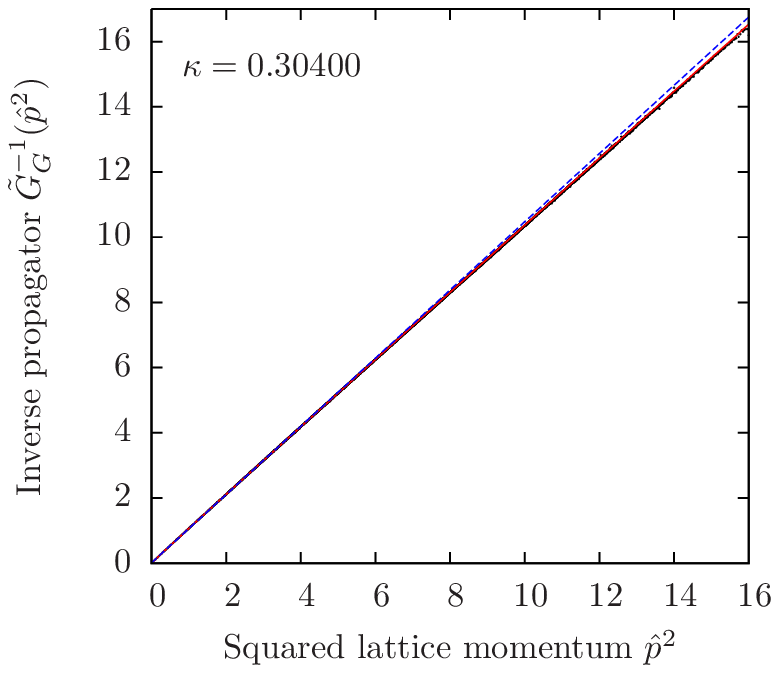}{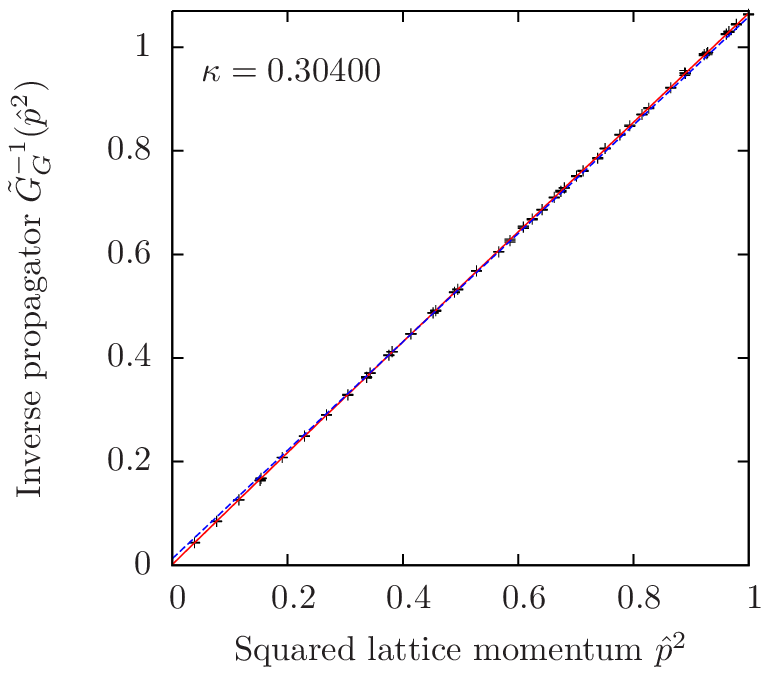}{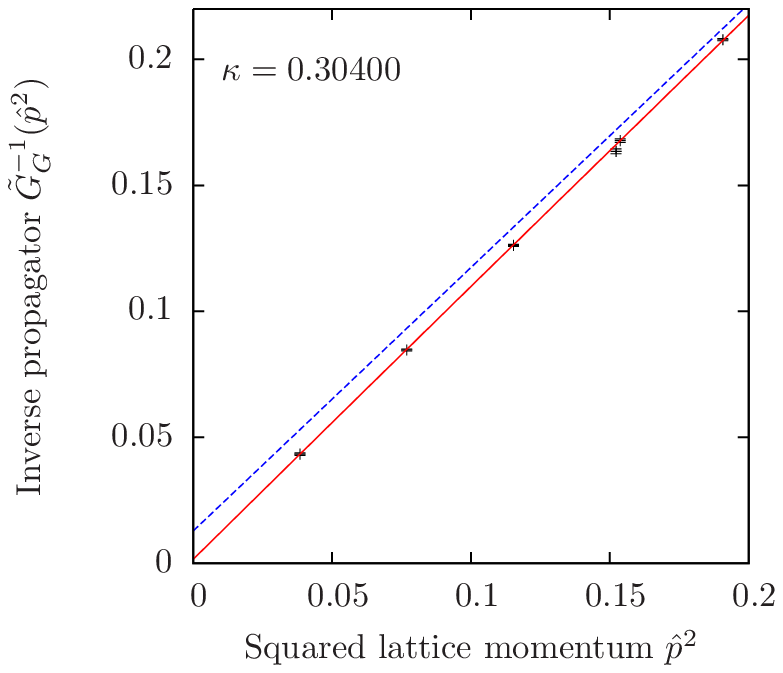}
{fig:GoldstonePropExampleArStrongCoup}
{The inverse lattice Goldstone propagators calculated in the Monte-Carlo runs specified in \tab{tab:Chap71EmployedRuns}
are presented versus the squared lattice momenta $\hat p^2$ together with the respective fits obtained from the 
fit approaches $f^{-1}_G(\hat p^2)$ in \eq{eq:GoldstoneFitAnsatz} (red solid line) and $l^{-1}_G(\hat p^2)$ 
in \eq{eq:GoldstoneFitAnsatzLinear} (blue dashed line) with $\gamma=4.0$. From left to right 
the three panel columns display the same data zooming in, however, on the vicinity of the origin at $\hat p^2 = 0$.
}
{Examples of Goldstone propagators at large quartic coupling constants.}
\includeTabNoHLines{|c|c|c|c|c|c|c|c|}{
\cline{3-8}
\multicolumn{2}{c|}{}& \multicolumn{3}{c|}{fit ansatz $f_G(\hat p^2)$} &  \multicolumn{3}{c|}{linear fit ansatz $l_G(\hat p^2)$}\\ \hline
$\kappa$             & $\gamma$   & $Z_{G}$     & $m_{G}$    &  $\chi^2/dof$  & $Z_{G}$   & $m_{G}$   &  $\chi^2/dof$ \\ \hline
$0.30039$     & $1.0$        & 0.9380(107) & 0.027(15)  &  1.00          & 0.9422(5) & 0.067(2)  & 2.61  \\
$0.30039$     & $2.0$        & 0.9431(52)  & 0.028(11)  &  0.81          & 0.9507(3) & 0.089(2)  & 4.79  \\
$0.30039$     & $4.0$        & 0.9457(27)  & 0.033(8)   &  0.94          & 0.9585(2) & 0.114(2)  & 6.19  \\\hline
$0.30400$     & $1.0$        & 0.9400(90)  & 0.029(10)  &  1.41          & 0.9403(4) & 0.066(1)  & 4.40  \\
$0.30400$     & $2.0$        & 0.9426(36)  & 0.032(7)   &  1.07          & 0.9476(2) & 0.084(1)  & 6.53  \\
$0.30400$     & $4.0$        & 0.9478(18)  & 0.038(4)   &  1.06          & 0.9559(1) & 0.111(1)  & 9.67  \\\hline
}
{tab:GoldstonePropExampleResultsAtStrongCoup}
{The results on the Goldstone renormalization factor $Z_G$ and the Goldstone mass $m_G$, obtained from
the fit approaches $f_G(\hat p^2)$ and $l_G(\hat p^2)$ as defined in \eq{eq:GoldstoneFitAnsatz} and 
\eq{eq:GoldstoneFitAnsatzLinear}, are listed for several settings of the parameter 
$\gamma$ together with the corresponding average squared residual per degree of freedom $\chi^2/dof$
associated to the respective fit. The underlying Goldstone lattice propagators have been calculated
in the Monte-Carlo runs specified in \tab{tab:Chap71EmployedRuns}.
}
{Comparison of the Goldstone propagator properties obtained from different extraction schemes at large quartic
coupling constants.}

In contrast to that the more elaborate fit ansatz $f_G(\hat p^2)$ yields much better values of $\chi^2/dof$ being close
to the expected value of one as can be seen in \tab{tab:GoldstonePropExampleResultsAtStrongCoup}. Moreover, the results on
the renormalization constant $Z_G$ and the Goldstone mass $m_G$ obtained from this ansatz remain consistent with respect to 
the specified errors when varying the constant $\gamma$. 
In the following the aforementioned quantities $Z_G$ and $m_G$ will therefore always be determined by means of the
here presented method based on the fit ansatz $f_G(\hat p^2)$ with a threshold value of $\gamma=4$, since this setting yields 
the most stable results, while still being consistent with the findings obtained at smaller values of $\gamma$.

\section{Analysis of the Higgs propagator}
\label{sec:HiggsPropAnalysisUpperBound}

Concerning the analysis of the Higgs propagator we will follow the same strategy as in the previous section.
Examples of the lattice Higgs propagator as obtained in the Monte-Carlo runs specified in \tab{tab:Chap71EmployedRuns} 
are presented in \fig{fig:HiggsPropExampleAtStrongCoup}. These numerical data have been fitted with the ansatz
\bea
\label{eq:HiggsPropFitAnsatz}
f^{-1}_H(\hat p^2) &=& \frac{\hat p^2 + \bar m_H^2
+ A\cdot \left[ 36 \left( \OneBosonLoopContEuc(\hat p^2, \bar m_H^2) - D_{H0}\right)  
+ 12 \left( \OneBosonLoopContEuc(\hat p^2, \bar m_G^2) - D_{G0}\right) \right] }{Z_0},  \quad\quad 
\eea
derived from renormalized perturbation theory in the Euclidean pure $\Phi^4$-theory at one-loop order.
The restriction to the pure $\Phi^4$-theory is again motivated by the same arguments already discussed in the preceding section.
In the above formula the one-loop contribution $\OneBosonLoopContEuc(p^2, \bar m^2)$ is defined as
\bea
\label{eq:DefOfContEuc1LoopBosContrib}
\OneBosonLoopContEuc(p^2, \bar m^2) &=& \frac{\arctanh\left( q \right)}{q}, \quad q = \sqrt{\frac{p^2}{4 \bar m^2 + p^2}}, 
\eea
the constants $D_{H0}$, $D_{G0}$ are given as
\bea
D_{H0} &=& \OneBosonLoopContEuc(0, \bar m_H^2) = 1, \\
D_{G0} &=& \OneBosonLoopContEuc(0, \bar m_G^2) = 1,
\eea
and $\bar m_H^2$, $A$, $Z_0$ are the free fit parameters. The parameter $\bar m_G^2$ is not treated 
as a free parameter here. Instead it is fixed to the value of $m_G$ resulting from the analysis of the Goldstone 
propagator by the method described in the previous section. The sole purpose of this approach is to achieve higher 
stability in the considered fit procedure, which otherwise would yield here only unsatisfactory results with respect 
to the associated statistical uncertainties.

Again, one can observe, however less clearly as compared to the previously discussed examples of the case of the Goldstone 
propagator, that the more elaborate fit ansatz $f_H(\hat p^2)$ describes the lattice data more accurately than the simple 
linear fit approach 
\beq
\label{eq:HiggsPropFitAnsatzLinear}
l^{-1}_H(\hat p^2) = \frac{\hat p^2 + m_H^2}{Z_H}.
\eeq
This is better observable in the lower row 
of \fig{fig:HiggsPropExampleAtStrongCoup} than in the upper row, where the differences tend to be rather negligible. The 
reason why the observed differences between the two fit approaches are less pronounced here, as compared to the situation
in the preceding section, simply is, that the threshold value $\gamma$ was chosen here to be $\gamma=1$ which will be motivated 
below. This setting of $\gamma$ is much smaller than the value underlying the previously discussed examples of the Goldstone 
propagators and causes the linear fit to come closer to the more elaborate ansatz $f_H(\hat p^2)$.

\includeFigTrippleDouble
{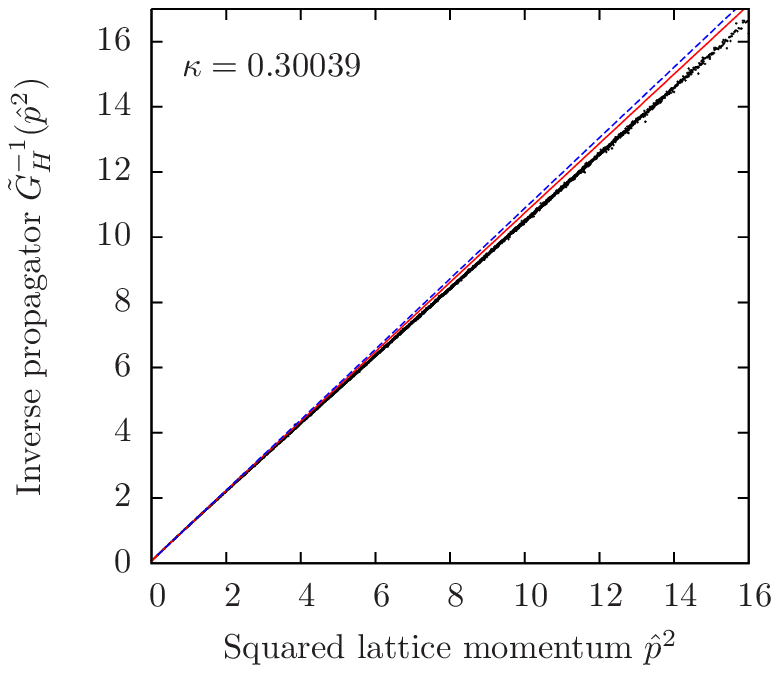}{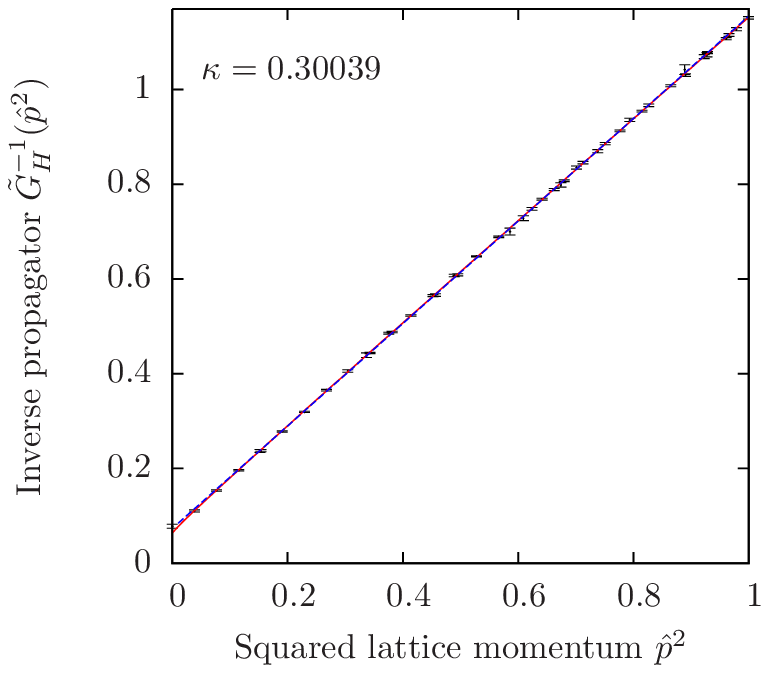}{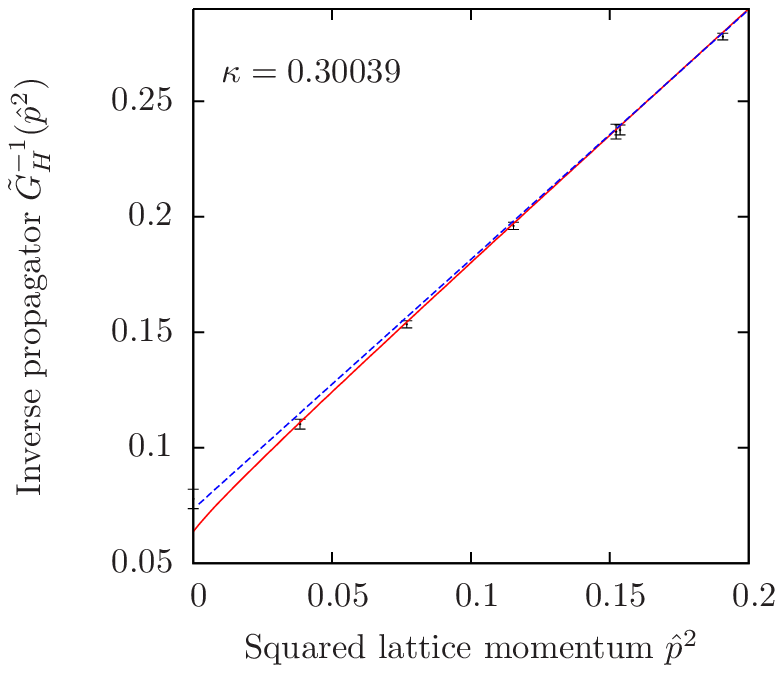}
{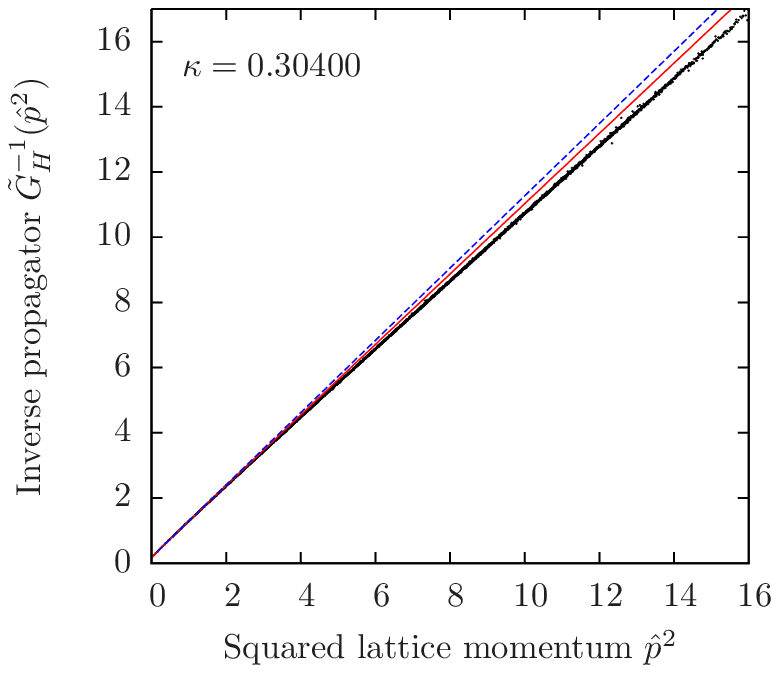}{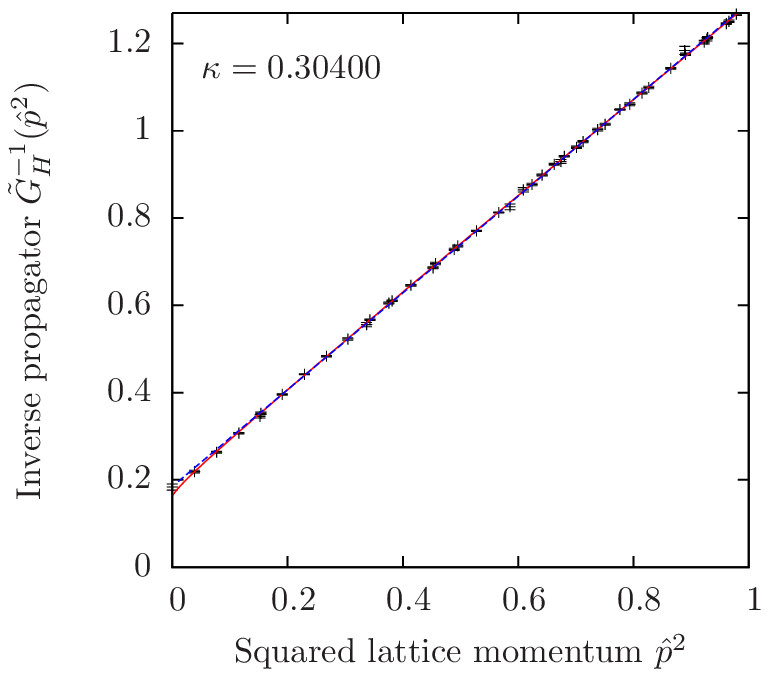}{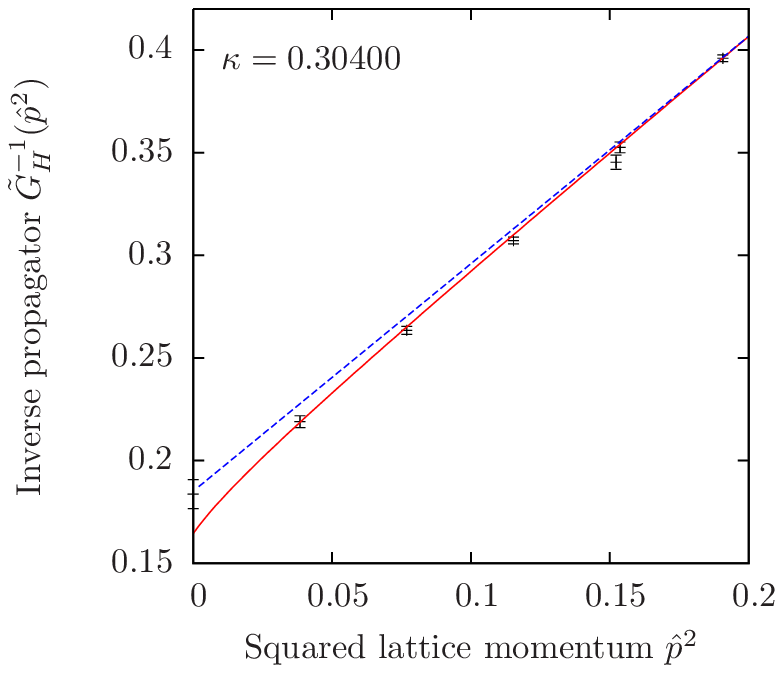}
{fig:HiggsPropExampleAtStrongCoup}
{The inverse lattice Higgs propagators calculated in the Monte-Carlo runs specified in \tab{tab:Chap71EmployedRuns}
are presented versus the squared lattice momenta $\hat p^2$ together with the respective fits obtained from the 
fit approaches $f^{-1}_H(\hat p^2)$ in \eq{eq:HiggsPropFitAnsatz} (red solid line) and $l^{-1}_H(\hat p^2)$ 
in \eq{eq:HiggsPropFitAnsatzLinear} (blue dashed line) with $\gamma=1.0$. From left to right the three 
panel columns display the same data zooming in, however, on the vicinity of the origin at $\hat p^2 = 0$.
}
{Examples of Higgs propagators at large quartic coupling constants.}

The Higgs propagator mass $m_{Hp}$ defined in \eq{eq:DefOfPropagatorMinkMass} and the Higgs pole mass $m_H$ together with its 
associated decay width $\Gamma_H$ given by the pole of the propagator on the second Riemann sheet according to 
\eq{eq:DefOfHiggsAndGoldstoneMassByPole} can then be obtained by defining the analytical continuation of the lattice 
propagator as $\tilde G_H^{c}(p_c) = f_H(p_c)$ and $\tilde G_H^{c}(p_c) = l_H(p_c)$, respectively. The results arising from 
the considered fit procedures are listed in \tab{tab:HiggsPropExampleResultsAtStrongCoup} for several values of the threshold
value $\gamma$. However, since the linear function $l_H(p_c)$ can not exhibit a branch cut structure, the pole mass equals the
propagator mass and the decay width is identical to zero when applying the linear fit approach. That is the reason why only 
the Higgs boson mass $m_H$ is presented in the latter scenario. We further remark that the values of $\Gamma_H$ arising along
with the determination of $m_H$ are -- as expected -- rather unstable due to the required analytical continuation of the 
propagator onto the second Riemann sheet. We therefore do not present these numbers here.

One observes in \tab{tab:HiggsPropExampleResultsAtStrongCoup} that the Higgs boson masses obtained from the linear fit ansatz
$l_H(\hat p^2)$ are again inconsistent with the respective results obtained at varying values of the threshold parameter $\gamma$,
thus rendering this latter approach unsuitable for the description of the Higgs propagator. This becomes also manifest through
the presented values of the average squared residual per degree of freedom $\chi^2/dof$ associated to the linear ansatz, which 
are clearly off the expected value of one (with the exception of the case of $\gamma=0.5$).

\includeTabNoHLines{|c|c|c|c|c|c|c|}{
\cline{3-7}
\multicolumn{2}{c|}{} &  \multicolumn{3}{c|}{Fit ansatz $f_H(\hat p^2)$} &  \multicolumn{2}{c|}{Fit ansatz $l_H(\hat p^2)$}  \\ \hline
$\kappa$ & $\gamma$   &  $m_{Hp}$  &  $m_{H}$       &  $\chi^2/dof$   &  $m_{H}$  &  $\chi^2/dof$    \\ \hline
0.30039  &  0.5       &  0.253(2)  & 0.296(83)      &  1.17           &  0.253(2)  & 1.13\\
0.30039  &  1.0       &  0.252(2)  & 0.253(2)       &  1.20           &  0.261(2)  & 1.62\\
0.30039  &  2.0       &  0.246(2)  & 0.249(2)       &  1.09           &  0.273(1)  & 2.58\\ \hline
0.30400  &  0.5       &  0.405(3)  & 0.406(3)       &  1.43           &  0.399(2)  & 1.75\\
0.30400  &  1.0       &  0.409(1)  & 0.410(1)       &  1.16           &  0.409(1)  & 2.23\\
0.30400  &  2.0       &  0.409(1)  & 0.412(1)       &  1.27           &  0.423(1)  & 4.63 \\ \hline
}
{tab:HiggsPropExampleResultsAtStrongCoup}
{The results on the Higgs propagator mass $m_{Hp}$ and the Higgs pole mass $m_H$
obtained from the fit approaches $f_H(\hat p^2)$ in \eq{eq:HiggsPropFitAnsatz}
and $l_H(\hat p^2)$ in \eq{eq:HiggsPropFitAnsatzLinear} are listed for several 
settings of the parameter $\gamma$ together with the corresponding average squared residual per degree 
of freedom $\chi^2/dof$ associated to the respective fit. For the linear fit ansatz $l_H(\hat p^2)$
only the pole mass is presented, since one finds $m_{Hp}\equiv m_H$ when constructing
the analytical continuation $\tilde G_H^c(p_c)$ through $l_H(p_c)$. The underlying 
Higgs lattice propagators have been calculated in the Monte-Carlo runs specified in \tab{tab:Chap71EmployedRuns}.
}
{Comparison of the Higgs propagator properties obtained from different extraction schemes at large quartic coupling
constants.}

Again the situation is very different in case of the more elaborate fit ansatz $f_H(\hat p^2)$ yielding significantly smaller values 
of $\chi^2/dof$. The presented results on the propagator mass $m_{Hp}$ as well as the pole mass $m_H$ are also in much better agreement
with the corresponding values obtained at varied threshold parameter $\gamma$. Moreover, the values of $m_{Hp}$ and $m_H$ are consistent 
with each other with respect to the given errors, finally justifying the identification of the Higgs boson mass with the propagator
mass $m_{Hp}$.

From the findings presented in \tab{tab:HiggsPropExampleResultsAtStrongCoup} one can conclude that selecting the threshold value
to be $\gamma=1$ for the analysis of the Higgs propagator is a very reasonable choice, which leads to consistent and satisfactory 
results. This is the setting that will be used for the subsequent investigation of the upper Higgs boson mass bounds to determine 
the propagator mass $m_{Hp}$. It is further remarked that the here chosen value of $\gamma$ is much 
smaller than the value $\gamma=4$ selected in the preceding section for the analysis of the Goldstone propagator. While this large 
setting worked well in the latter scenario, it leads to less consistent results in the here considered case and has therefore been excluded 
from the given presentation.

\section{Dependence of the Higgs boson mass on the bare coupling constant $\lambda$}
\label{sec:DepOfHiggsMassonLargeLam}

We now turn to the question whether the largest Higgs boson mass is indeed obtained at infinite bare quartic coupling constant 
for a given set of quark masses and a given cutoff $\Lambda$ as one would expect from perturbation theory. Since perturbation 
theory may not be trustworthy in the regime of large bare coupling constants, the actual dependence of the Higgs boson 
mass on the bare quartic coupling constant $\lambda$ in the scenario of strong interactions is explicitly checked here by means 
of direct Monte-Carlo calculations. The final answer of what bare coupling constant produces the largest Higgs boson mass will then 
be taken as input for the investigation of the upper mass bound in the subsequent section.

For this purpose some numerical results on the Higgs propagator mass $m_{Hp}$ are plotted versus the bare quartic coupling constant 
$\lambda$ in \fig{fig:StrongCoulingHiggsMassDepOnQuartCoup}a. The presented data have been obtained for a cutoff that was intended to 
be kept constant at approximately $\Lambda\approx \GEV{1540}$ by an appropriate tuning of the hopping parameter, while the degenerate 
Yukawa coupling constants were fixed according to the tree-level relation in \eq{eq:treeLevelTopMass} aiming at the reproduction of 
the top quark mass. One clearly observes that the Higgs boson mass monotonically rises with increasing values of the bare coupling 
constant $\lambda$ until it finally converges to the $\lambda=\infty$ result, which is depicted by the horizontal line in the presented 
plot. From this result one can conclude that the largest Higgs boson mass is indeed obtained at infinite bare quartic coupling constant,
as expected. The forthcoming search for the upper mass bound will therefore be restricted to the scenario of $\lambda=\infty$.

\includeFigDouble{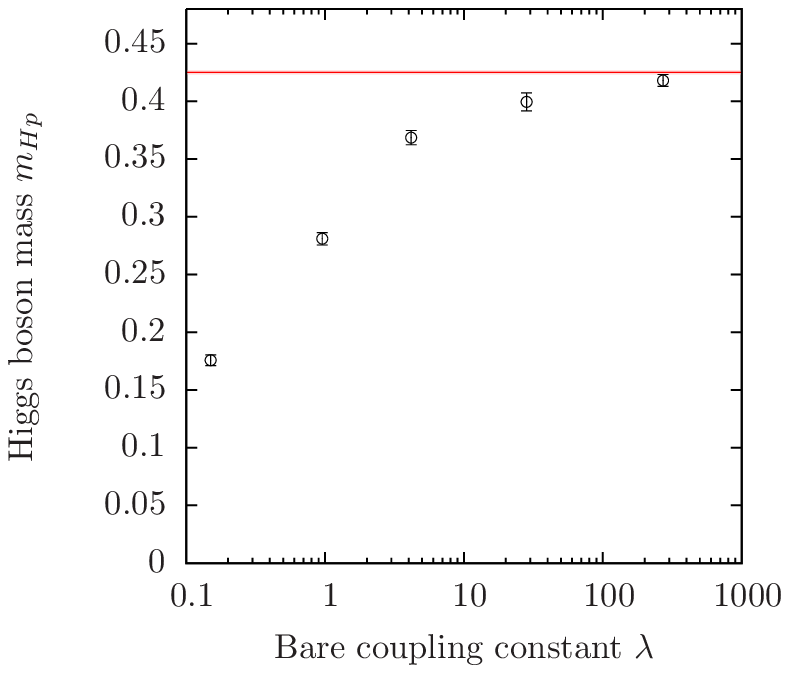}{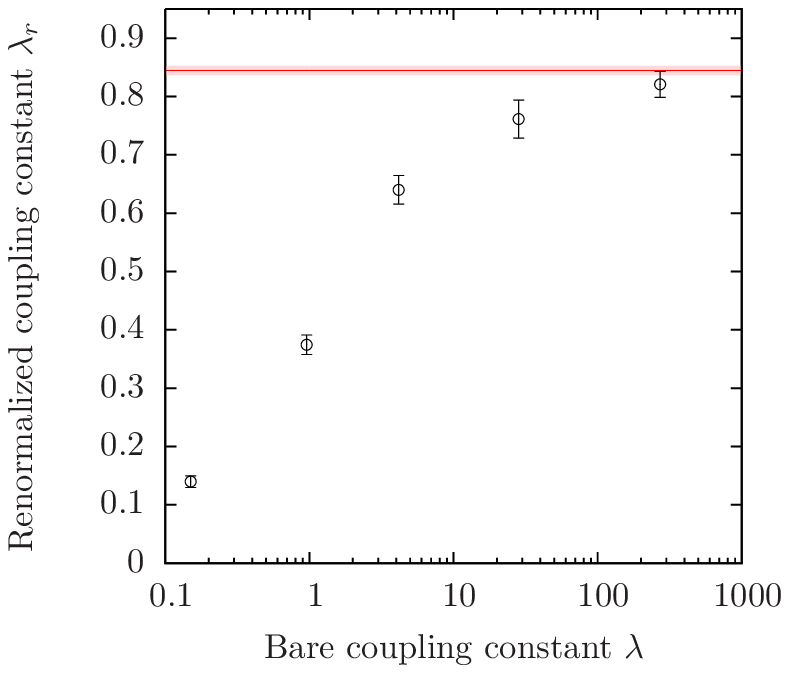}
{fig:StrongCoulingHiggsMassDepOnQuartCoup}
{The Higgs boson mass $m_{Hp}$ and the renormalized quartic coupling constant $\lambda_r$ are shown versus the bare coupling
constant $\lambda$ in panels (a) and (b), respectively. These results have been obtained in direct Monte-Carlo calculations 
on a \lattice{16}{32} with the degenerate Yukawa coupling constants fixed according to the tree-level relation in 
\eq{eq:treeLevelTopMass} aiming at the reproduction of the top quark mass. The hopping parameter was tuned with the intention
to hold the cutoff constant, while the actually obtained values of $\Lambda$ fluctuate here between $\GEV{1504}$ and
$\GEV{1549}$. The horizontal lines depict the corresponding results at infinite bare coupling constant $\lambda=\infty$, and
the highlighted bands mark the associated statistical uncertainties.
}
{Dependence of the Higgs boson mass $m_{Hp}$ and the renormalized quartic coupling constant $\lambda_r$ on the 
bare quartic coupling constant $\lambda$.}

Furthermore, the corresponding behaviour of the renormalized quartic coupling constant $\lambda_r$ as defined in \eq{eq:DefOfRenQuartCoupling} 
is presented in \fig{fig:StrongCoulingHiggsMassDepOnQuartCoup}b. As expected one observes a monotonically rising dependence
of $\lambda_r$ on the bare coupling constant $\lambda$, eventually converging to the $\lambda=\infty$ result depicted by the 
horizontal line.

\section{Results on the upper Higgs boson mass bound}
\label{chap:ResOnUpperBound}
 
We now turn to the actually intended non-perturbative determination of the cutoff-dependent upper Higgs boson mass bound 
$m_H^{up}(\Lambda)$. Given the knowledge about the dependence of the Higgs boson mass on the bare quartic self-coupling 
constant $\lambda$ the search for the desired upper mass bound can safely be restricted to the scenario of an infinite 
bare quartic coupling constant, \ie $\lambda=\infty$. Moreover, we will restrict the investigation here to the mass 
degenerate case with $y_t=y_b$, since the fermion determinant $\det(\fermiMat)$ can be proven to be real 
in this scenario as discussed in \sect{sec:modelDefinition}.

Concerning the cutoff parameters $\Lambda$ that are reachable with the intended lattice calculations, a couple of 
restrictions limit the range of the accessible energy scales. On the one hand all particle masses have to be small compared to 
$\Lambda$ to avoid unacceptably large cutoff effects, on the other hand all masses have to be large compared to the inverse 
lattice side lengths to bring the finite volume effects to a tolerable level. As a minimal requirement we demand here that 
all particle masses $\hat m\in\{m_t, m_b, m_H \}$ in lattice units fulfill 
\bea
\label{eq:RequirementsForLatMass}
\hat m < 0.5& \quad \mbox{and} \quad & \hat m\cdot L_{s,t}>2,
\eea
which already is a rather loose condition in comparison with the common situation in QCD, where one usually demands
at least $\hat m \cdot L_{s,t}>3$. In this model, however, the presence of massless Goldstone modes is known to induce
{\it algebraic} finite size effects, which is why it is not meaningful to impose a much stronger constraint in 
\eq{eq:RequirementsForLatMass}, since the quantity $\hat m\cdot L_{s,t}$ only controls the strength of the exponentially 
suppressed finite size effects caused by the massive particles.

Employing a top mass of $\GEV{175}$ and a Higgs boson mass of roughly $\GEV{700}$, which will turn out to be justified 
after the upper mass bound has eventually been established, it should therefore be feasible to reach energy scales between 
$\GEV{1400}$ and $\GEV{2800}$ on a \latticeX{32}{32}{.}

For the purpose of investigating the cutoff dependence of the upper mass bound a series of direct Monte-Carlo
calculations has been performed with varying hopping parameters $\kappa$ associated to cutoffs covering approximately
the given range of reachable energy scales. At each value of $\kappa$ the Monte-Carlo computation has been
rerun on several lattice sizes to examine the respective strength of the finite volume effects, ultimately allowing for
the infinite volume extrapolation of the obtained lattice results. In addition, a corresponding series of Monte-Carlo
calculations has been performed in the pure $\Phi^4$-theory, which will finally allow to address the question for the
fermionic contributions to the upper Higgs boson mass bound. The model parameters underlying these two series of lattice 
calculations are presented in \tab{tab:SummaryOfParametersForUpperHiggsMassBoundRuns}.

\includeTab{|ccccccccc|}
{
$\kappa$ & $L_s$                    & $L_t$ & $N_f$ &  $\hat \lambda$ & $\hat y_t$     & $\hat y_b/\hat y_t$ & $1/v$ & $\Lambda$ \\
\hline
0.30039  & 12,16,20,24,32  &   32  &  1    &  $\infty$         & 0.55139        & 1          & $\approx 7.7$   & $\GEV{\approx 2370 }$\\
0.30148  & 12,16,20,24,32  &   32  &  1    &  $\infty$         & 0.55239        & 1          & $\approx 6.5$   & $\GEV{\approx 1990 }$\\
0.30274  & 12,16,20,24,32  &   32  &  1    &  $\infty$         & 0.55354        & 1          & $\approx 5.6$   & $\GEV{\approx 1730 }$\\
0.30400  & 12,16,20,24,32  &   32  &  1    &  $\infty$         & 0.55470        & 1          & $\approx 5.0$   & $\GEV{\approx 1550 }$\\ \hline
0.30570  & 12,16,20,24,32  &   32  &  1    &  $\infty$         & 0              & --         & $\approx 9.0$   & $\GEV{\approx 2810 }$\\
0.30680  & 12,16,20,24,32  &   32  &  1    &  $\infty$         & 0              & --         & $\approx 7.1$   & $\GEV{\approx 2220 }$\\
0.30780  & 12,16,20,24,32  &   32  &  1    &  $\infty$         & 0              & --         & $\approx 6.2$   & $\GEV{\approx 1910 }$\\
0.30890  & 12,16,20,24,32  &   32  &  1    &  $\infty$         & 0              & --         & $\approx 5.5$   & $\GEV{\approx 1700 }$\\
0.31040  & 12,16,20,24,32  &   32  &  1    &  $\infty$         & 0              & --         & $\approx 4.9$   & $\GEV{\approx 1500 }$\\
}
{tab:SummaryOfParametersForUpperHiggsMassBoundRuns}
{The model parameters of the Monte-Carlo runs underlying the subsequent lattice calculation of the upper Higgs boson mass bound
are presented. In total, a number of 45 Monte-Carlo runs have been performed for that purpose. The available statistics of 
generated field configurations $\Nconf$ varies depending on the respective lattice volume. In detail we have $\Nconf\approx 20,000$ 
for $12\le L_s\le 16$, $\Nconf\approx10,000-15,000$ for $L_s = 20$, $\Nconf\approx8,000-16,000$ for $L_s=24$, 
$\Nconf\approx 3,000-5,000$ for $L_s=32$.
The numerically determined values of $1/v$ and $\Lambda$ are also approximately given. These numbers vary, of course, depending
on the respective lattice volumes and serve here only for the purpose of a rough orientation. The degenerate Yukawa coupling 
constants in the upper four rows have been chosen according to the tree-level relation in \eq{eq:treeLevelTopMass} aiming at the 
reproduction of the phenomenologically known top quark mass. In the other cases it is exactly set to zero recovering the pure 
$\Phi^4$-theory.}
{Model parameters of the Monte-Carlo runs underlying the lattice calculation of the upper Higgs boson mass bound.}

However, before discussing the obtained lattice results, it is worthwhile to recall what behaviour of the considered
observables is to be expected from the knowledge of earlier lattice investigations. For the pure $\Phi^4$-theory and neglecting 
any double-logarithmic contributions the cutoff dependence of the Higgs boson mass as well as the renormalized quartic coupling 
constant has been found in \Refs{Luscher:1987ay,Luscher:1987ek,Luscher:1988uq} to be of the form
\bea
\label{eq:StrongCouplingLambdaScalingBeaviourMass}
\frac{m_{Hp}}{a} &=& A_m \cdot \left[\log(\Lambda^2/\mu^2) + B_m \right]^{-1/2}, \\
\label{eq:StrongCouplingLambdaScalingBeaviourLamCoupling}
\lambda_r &=& A_\lambda \cdot \left[\log(\Lambda^2/\mu^2) + B_\lambda \right]^{-1},
\eea
where $\mu$ denotes some unspecified scale, and $A_{m,\lambda}\equiv A_{m,\lambda}(\mu)$, $B_{m,\lambda}\equiv B_{m,\lambda}(\mu)$ 
are constants. Since this result has been established in the pure $\Phi^4$-theory, it is thus worthwhile to ask 
whether these scaling laws still hold in the considered Higgs-Yukawa model including the
coupling to the fermions. In that respect it is remarked that the same functional dependence has 
also been observed in other analytical studies, for instance in \Ref{Fodor:2007fn} examining 
a Higgs-Yukawa model in continuous Euclidean space-time based, however, on an one-component Higgs field. 
In that study the running of the renormalized coupling constants with varying cutoff has been 
investigated by means of renormalized perturbation theory in the large $N_f$-limit. Furthermore, 
the scaling behaviour of the renormalized Yukawa coupling constant has also been derived. It was 
found to be
\bea
\label{eq:StrongCouplingLambdaScalingBeaviourYCoupling}
y_r &=& A_y \cdot \left[\log(\Lambda^2/\mu^2) + B_y \right]^{-1/2},
\eea
where $A_y\equiv A_y(\mu)$ and $B_y\equiv B_y(\mu)$ are again so far unspecified constants and $y_r$ stands 
here for the renormalized top and bottom Yukawa coupling constants $y_{t,r}$ and $y_{b,r}$, respectively, 
as defined in \eq{eq:DefOfRenYukawaConst}.

Now, the numerically obtained Higgs boson masses $m_{Hp}$ resulting in the above specified lattice calculations are finally
presented in \fig{fig:UpperHiggsCorrelatorBoundFiniteVol}, where panel (a) refers to the full Higgs-Yukawa model while panel (b)
displays the corresponding results in the pure $\Phi^4$-theory. To illustrate the influence of the finite lattice volume 
those results, belonging to the same parameter sets, differing only in the underlying lattice size, are connected by dotted 
lines to guide the eye. From these findings one learns that the model indeed exhibits strong finite volume effects when 
approaching the upper limit of the defined interval of reachable cutoffs, as expected. 

\includeFigDouble{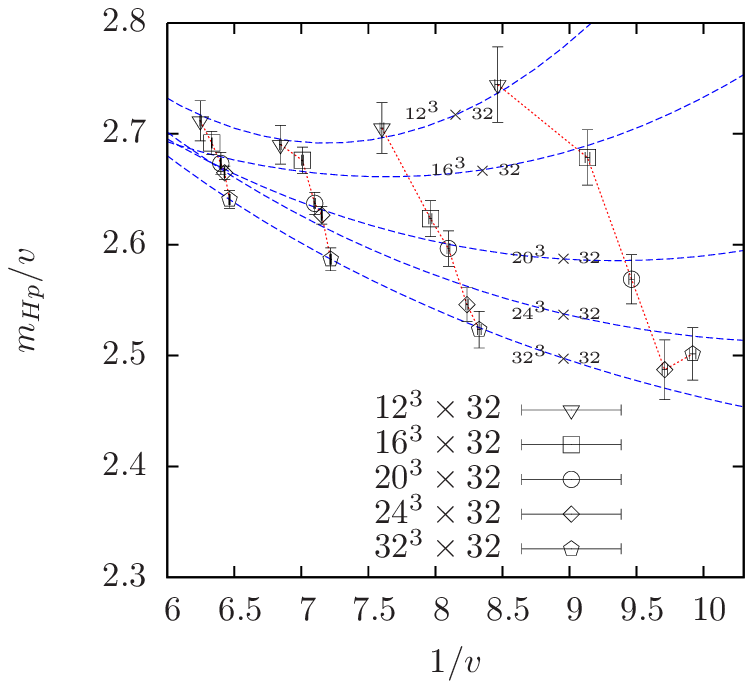}{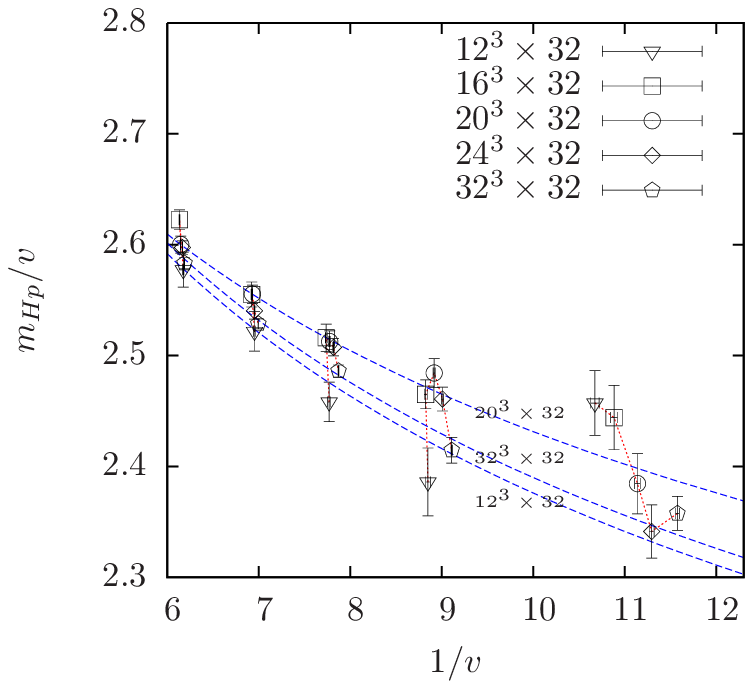}
{fig:UpperHiggsCorrelatorBoundFiniteVol}
{The Higgs propagator mass $m_{Hp}$ is presented in units of the vacuum expectation value $v$ 
versus $1/v$. These results have been determined in the direct Monte-Carlo calculations specified in
\tab{tab:SummaryOfParametersForUpperHiggsMassBoundRuns}. Those runs with identical parameter sets differing 
only in the underlying lattice volume are connected via dotted lines to illustrate the effects of the 
finite volume. The dashed curves depict the fits of the lattice results according to the finite size expectation
in \eq{eq:RenormHiggsMassFitFormula} as explained in the main text. Panel (a) refers to the full Higgs-Yukawa model,
while panel (b) shows the corresponding results of the pure $\Phi^4$-theory.
}
{Dependence of the Higgs propagator mass on $1/v$ at infinite bare quartic coupling constant.}

In \fig{fig:UpperHiggsCorrelatorBoundFiniteVol}a one sees that the Higgs boson mass seems to increase with
the cutoff $\Lambda$ on the smaller lattice sizes. This, however, is only a finite volume effect. On the larger 
lattices the Higgs boson mass decreases with growing $\Lambda$ as expected from the triviality property of the 
Higgs sector. In comparison to the results obtained in the pure $\Phi^4$-theory shown in \fig{fig:UpperHiggsCorrelatorBoundFiniteVol}b
the aforementioned finite size effects, being of order $\proz{10}$ here, are much stronger and can thus be 
ascribed to the influence of the coupling to the fermions. This effect directly arises from the top quark 
being the lightest physical particle in the here considered scenario.

At this point it is worthwhile to ask whether the observed finite volume effects can also be understood by some quantitative consideration.
For the weakly interacting regime this could be achieved, for instance, by computing the constraint effective 
potential~\cite{Fukuda:1974ey,O'Raifeartaigh:1986hi} (CEP) in terms of the bare model parameters for a given finite volume 
as discussed in \Ref{Gerhold:2009ub}, which then allowed to predict the numerical lattice data for given bare model parameters.
In contrast to that scenario the same calculation is not directly useful in the present situation, since the underlying (bare)
perturbative expansion would break down due to the bare quartic coupling constant being infinite here. This problem 
can be cured by parametrizing the four-point interaction in terms of the renormalized quartic coupling constant. Starting from the 
definition of $\lambda_r$ in \eq{eq:DefOfRenQuartCoupling} one can directly derive an estimate for the Higgs boson mass in 
terms of $\lambda_r$ according to
\bea
m_{He}^2 &=& 8\lambda_r v^2 
-\frac{1}{v} \derive{}{\breve v} U_F[\breve v]\Bigg|_{\breve v = v} + \derive{^2}{\breve v^2} 
U_F[\breve v]\Bigg|_{\breve v = v} \\
\label{eq:DefOfFermionicContToEffPot}
U_{F}[\breve v] &=&
\frac{-2N_f}{L_s^3\cdot L_t}\cdot \sum\limits_{p} \log\left|\nu^+(p) + y_t \breve v \left(1-\frac{1}{2\rho}\nu^+(p)\right) 
\right|^2\nonumber\\
&+& \frac{-2N_f}{L_s^3\cdot L_t}\cdot \sum\limits_{p}  \log\left|\nu^+(p) + y_b \breve v \left(1-\frac{1}{2\rho}\nu^+(p)\right)  \right|^2. 
\label{eq:FermionEffectivePot}
\eea
which respects all contributions of order $O(\lambda_r)$. It is remarked that the above contribution $U_F[\breve v]$ contains all fermionic loops 
in the background of a constant scalar field and has already been discussed in \Ref{Gerhold:2009ub}, while the underlying definition of
the eigenvalues $\nu^\pm(p)$ of the free overlap operator with $\pm \IM(\nu^\pm(p))\ge 0$ has been taken from \Ref{Gerhold:2007yb}.

Combining the above result with the expected scaling law given in \eq{eq:StrongCouplingLambdaScalingBeaviourLamCoupling} a crude estimate for the 
observed behaviour of the Higgs boson mass presented in \fig{fig:UpperHiggsCorrelatorBoundFiniteVol} can be established according to
\bea
\label{eq:RenormHiggsMassFitFormula}
m_{He}^2 &=& \frac{8v^2 A'_\lambda}{\log(v^{-2}) + B'_\lambda}  -\frac{1}{v} \derive{}{\breve v} U_F[\breve v]\Bigg|_{\breve v = v} + \derive{^2}{\breve v^2} 
U_F[\breve v]\Bigg|_{\breve v = v}, 
\eea
where double-logarithmic terms have been neglected and $A'_\lambda$, $B'_\lambda$ are so far unspecified parameters.

Since the value of the renormalized quartic coupling constant $\lambda_r$ is not known apriori, the idea is here to use the result in 
\eq{eq:RenormHiggsMassFitFormula} as a fit ansatz with the free fit parameters $A'_\lambda$ and $B'_\lambda$ to fit the observed 
finite volume behaviour of the Higgs boson mass presented in \fig{fig:UpperHiggsCorrelatorBoundFiniteVol}. These lattice data
have been given in units of the vacuum expectation value $v$, plotted versus $1/v$, to allow for the intended
direct comparison with the analytical finite volume expression in \eq{eq:RenormHiggsMassFitFormula}. The resulting 
fits are depicted by the dashed curves in \fig{fig:UpperHiggsCorrelatorBoundFiniteVol}, where the free parameters $A'_\lambda$ and $B'_\lambda$
have independently been adjusted for every presented series of constant lattice volume in the full Higgs-Yukawa model and the pure 
$\Phi^4$-theory, respectively. Applying the above fit ansatz simultaneously to all available data does not lead to satisfactory results, 
since the renormalized quartic coupling constant itself also depends significantly on the underlying lattice volume, as will be 
seen later in this section. 

One can then observe in \fig{fig:UpperHiggsCorrelatorBoundFiniteVol} that this fit approach can describe the actual finite
volume cutoff dependence of the presented Higgs boson mass satisfactorily well, unless the vacuum expectation value $v$ becomes
too small. In that case the model does no longer exhibit the expected (infinite volume) critical behaviour in 
\eqs{eq:StrongCouplingLambdaScalingBeaviourMass}{eq:StrongCouplingLambdaScalingBeaviourLamCoupling} which the derivation 
of the above fit ansatz was built upon. Staying away from that regime, however, the observed finite volume behaviour of 
the Higgs boson mass can be well understood by means of the analytical expression
in \eq{eq:RenormHiggsMassFitFormula}.

To obtain the desired upper Higgs boson mass bounds $m_H^{up}(\Lambda)$ these finite volume results have to be extrapolated 
to the infinite volume limit and the renormalization factor $Z_G$ has to be properly considered. For that purpose the finite 
volume dependence of the Monte-Carlo results on the renormalized vev $v_r=v/\sqrt{Z_G}$ and the Higgs boson mass $m_{Hp}$, 
as obtained for the two scenarios of the full Higgs-Yukawa model and the pure $\Phi^4$-theory, is explicitly shown in 
\fig{fig:FiniteVolumeEffectsOfUpperHiggsMassBoundVEVandMH}. One sees in these plots that the finite volume effects are 
rather mild at the largest investigated hopping parameters $\kappa$ corresponding to the lowest considered values of 
the cutoff $\Lambda$, while the renormalized vev as well as the Higgs boson mass itself vary strongly with increasing 
lattice size $L_s$ at the smaller presented hopping parameters, as expected.

It is well known from lattice investigations of the pure $\Phi^4$-theory~\cite{Hasenfratz:1989ux,Hasenfratz:1990fu,Gockeler:1991ty}
that the vev as well as the mass receive strong contributions from the Goldstone modes, inducing finite volume effects 
of algebraic form starting at order $O(L_s^{-2})$. The next non-trivial finite volume contribution was shown to 
be of order $O(L_s^{-4})$. In \fig{fig:FiniteVolumeEffectsOfUpperHiggsMassBoundVEVandMH} the obtained data are 
therefore plotted versus $1/L_s^2$. Moreover, the aforementioned observation justifies to apply the linear fit ansatz 
\beq
\label{eq:LinFit}
f^{(l)}_{v,m}(L_s^{-2}) = A^{(l)}_{v,m} + B^{(l)}_{v,m}\cdot L_s^{-2}
\eeq
to extrapolate these data to the infinite volume limit, where the free fitting parameters $A^{(l)}_{v,m}$ and 
$B^{(l)}_{v,m}$ with the subscripts $v$ and $m$ refer to the renormalized vev $v_r$ and the Higgs boson mass 
$m_{Hp}$, respectively.

\includeFigDoubleDoubleHere{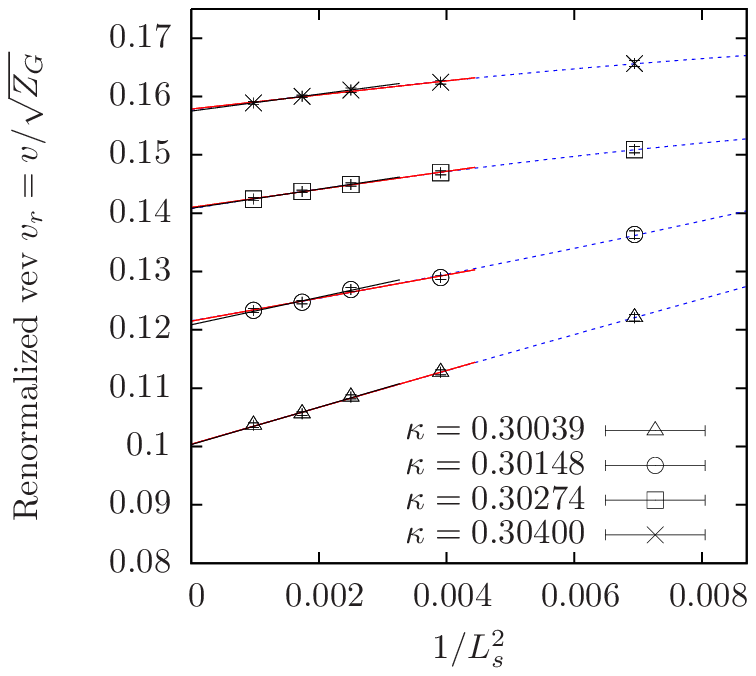}{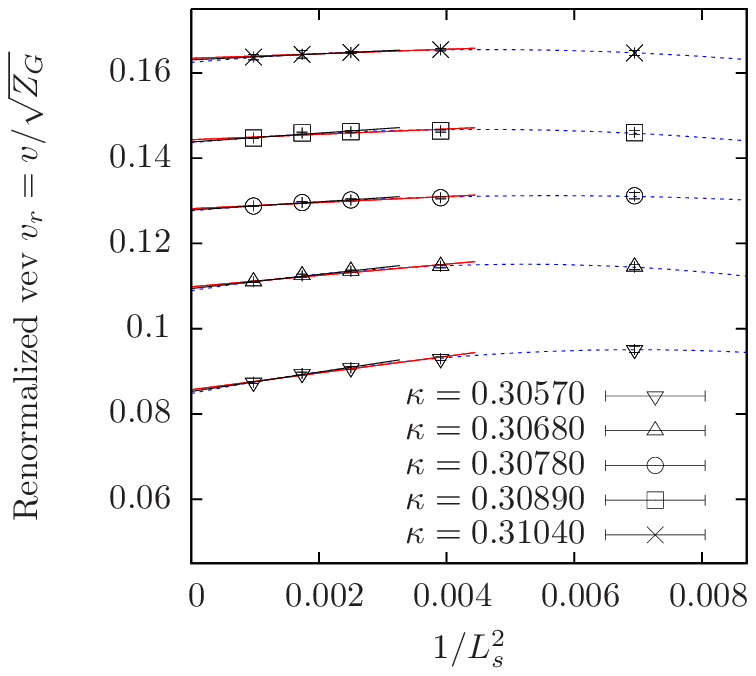}
{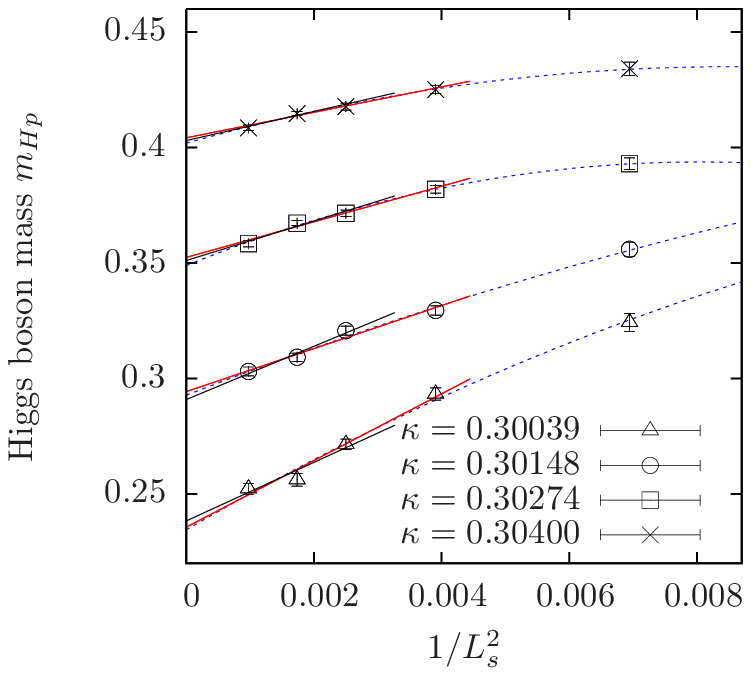}{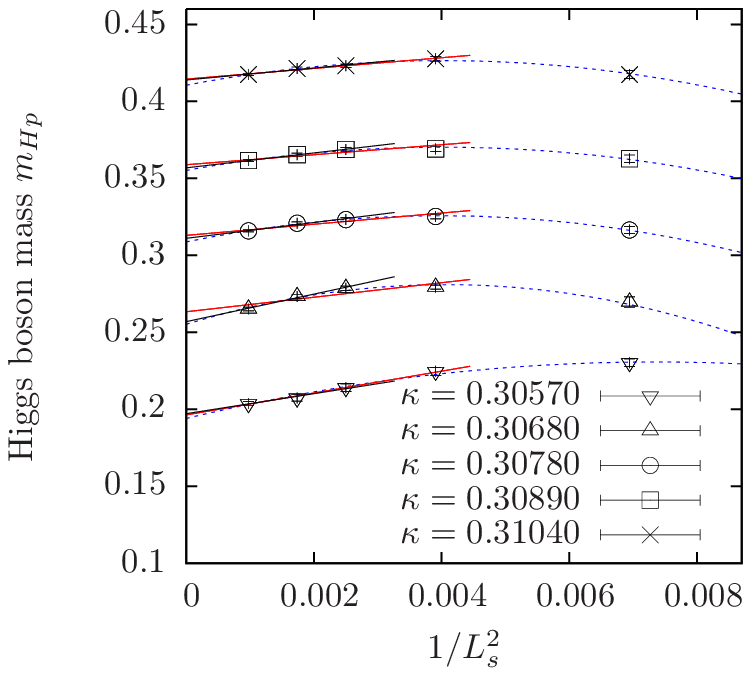}
{fig:FiniteVolumeEffectsOfUpperHiggsMassBoundVEVandMH}
{The dependence of the renormalized vev $v_r = v/\sqrt{Z_G}$ and the Higgs propagator mass $m_{Hp}$
on the squared inverse lattice side length $1/L^2_s$ is presented in the upper and the lower panel rows, respectively, as
determined in the direct Monte-Carlo calculations specified in \tab{tab:SummaryOfParametersForUpperHiggsMassBoundRuns}.
Panels (a) and (c) show the results for the full Higgs-Yukawa model, while panels (b) and (d) refer to the pure $\Phi^4$-theory.
In all plots the dashed curves display the parabolic fits according to the fit ansatz in 
\eq{eq:ParaFit}, while the solid lines depict the linear fits resulting from \eq{eq:LinFit} for the two lower threshold 
values $L_s'=16$ (red) and $L_s'=20$ (black).
}
{Dependence of the renormalized vev $v_r = v/\sqrt{Z_G}$ and the Higgs propagator mass $m_{Hp}$
on the squared inverse lattice side length $1/L^2_s$ at infinite bare quartic coupling constant.}

\includeTab{|c|c|c|c|c|}{
\multicolumn{5}{|c|}{Vacuum expectation value $v$} \\ \hline
$\kappa$  	 & $A^{(l)}_v$, $L'_s=16$  & $A^{(l)}_v$, $L'_s=20$         & $A^{(p)}_v$         & $v_r$               \\ \hline
$\,0.30039\,$  	 & $\, 0.1004(3) \, $   & $\, 0.1003(6) \, $  & $\, 0.1004(5) \, $  & $\, 0.1004(5)(1)$  \\ 
$\,0.30148\,$  	 & $\, 0.1215(5) \, $   & $\, 0.1209(6) \, $  & $\, 0.1216(8) \, $  & $\, 0.1213(6)(4)$  \\ 
$\,0.30274\,$  	 & $\, 0.1410(1) \, $   & $\, 0.1408(1) \, $  & $\, 0.1408(1) \, $  & $\, 0.1409(1)(1)$  \\ 
$\,0.30400\,$  	 & $\, 0.1579(2) \, $   & $\, 0.1575(1) \, $  & $\, 0.1576(2) \, $  & $\, 0.1577(2)(2)$  \\ 
$\,0.30570\,$  	 & $\, 0.0857(4) \, $   & $\, 0.0852(4) \, $  & $\, 0.0848(2) \, $  & $\, 0.0852(3)(5)$  \\ 
$\,0.30680\,$  	 & $\, 0.1099(4) \, $   & $\, 0.1094(2) \, $  & $\, 0.1089(1) \, $  & $\, 0.1097(3)(5)$  \\ 
$\,0.30780\,$  	 & $\, 0.1282(3) \, $   & $\, 0.1278(1) \, $  & $\, 0.1277(2) \, $  & $\, 0.1279(2)(3)$  \\ 
$\,0.30890\,$  	 & $\, 0.1443(5) \, $   & $\, 0.1438(5) \, $  & $\, 0.1436(4) \, $  & $\, 0.1439(5)(4)$  \\ 
$\,0.31040\,$  	 & $\, 0.1634(2) \, $   & $\, 0.1630(1) \, $  & $\, 0.1625(2) \, $  & $\, 0.1630(2)(5)$  \\  \hline
\multicolumn{5}{|c|}{Higgs propagator mass $m_{Hp}$} \\ \hline
$\kappa$  	 & $A^{(l)}_m$, $L'_s=16$  & $A^{(l)}_m$, $L'_s=20$         & $A^{(p)}_m$         & $m_{Hp}$               \\ \hline
$\,0.30039\,$  	 & $\, 0.2356(41)\, $   & $\, 0.2382(70)\, $  & $\, 0.2344(67)\, $  & $\, 0.2361(61)(19)$  \\ 
$\,0.30148\,$  	 & $\, 0.2943(29)\, $   & $\, 0.2908(39)\, $  & $\, 0.2928(40)\, $  & $\, 0.2926(36)(18)$  \\ 
$\,0.30274\,$  	 & $\, 0.3524(20)\, $   & $\, 0.3510(38)\, $  & $\, 0.3489(23)\, $  & $\, 0.3508(28)(18)$  \\ 
$\,0.30400\,$  	 & $\, 0.4042(14)\, $   & $\, 0.4030(25)\, $  & $\, 0.4018(15)\, $  & $\, 0.4030(19)(12)$  \\ 
$\,0.30570\,$  	 & $\, 0.1964(10)\, $   & $\, 0.1971(16)\, $  & $\, 0.1940(25)\, $  & $\, 0.1958(18)(16)$  \\ 
$\,0.30680\,$  	 & $\, 0.2633(42)\, $   & $\, 0.2568(20)\, $  & $\, 0.2552(30)\, $  & $\, 0.2584(32)(41)$  \\ 
$\,0.30780\,$  	 & $\, 0.3130(17)\, $   & $\, 0.3110(14)\, $  & $\, 0.3087(7) \, $  & $\, 0.3109(13)(22)$  \\ 
$\,0.30890\,$  	 & $\, 0.3589(17)\, $   & $\, 0.3568(3) \, $  & $\, 0.3552(10)\, $  & $\, 0.3570(12)(19)$  \\ 
$\,0.31040\,$  	 & $\, 0.4145(8) \, $   & $\, 0.4139(14)\, $  & $\, 0.4105(15)\, $  & $\, 0.4130(13)(20)$  \\ 
}
{tab:ResultOfUpperHiggsMassFiniteVolExtrapolation1}
{The results of the infinite volume extrapolations of the Monte-Carlo data of the renormalized vev $v_r$ and the Higgs 
boson mass $m_{Hp}$ are presented as obtained from the parabolic ansatz in \eq{eq:ParaFit} and the linear approach in 
\eq{eq:LinFit} for the considered lower threshold values $L_s'=16$ and $L_s'=20$. 
The final results on $v_r$ and $m_{Hp}$, displayed in the very right column, are determined here by averaging over
the parabolic and the two linear fit approaches. An additional, systematic uncertainty of these final results is specified
in the second pair of brackets taken from the largest observed deviation among all respective fit results.}
{Infinite volume extrapolation of the Monte-Carlo data for the renormalized vev $v_r$ and the Higgs 
boson mass $m_{Hp}$ at infinite bare quartic coupling constant.}

To take the presence of higher order terms in $1/L_{s}^{2}$ into account only the largest lattice sizes are included into 
this linear fit. Here, we select all lattice volumes with $L_s\ge L'_s$. As a consistency check, testing the dependence 
of the resulting infinite volume extrapolations on the choice of the fit procedure, the lower threshold value $L'_s$ is 
varied. The respective results are listed in \tab{tab:ResultOfUpperHiggsMassFiniteVolExtrapolation1}.
Moreover, the parabolic fit ansatz 
\beq
\label{eq:ParaFit}
f^{(p)}_{v,m}(L_s^{-2}) = A^{(p)}_{v,m} + B^{(p)}_{v,m}\cdot L_s^{-2} + C^{(p)}_{v,m}\cdot L_s^{-4}
\eeq
is additionally considered. It is applied to the whole range of available lattice sizes. The deviations
between the various fitting procedures with respect to the resulting infinite volume extrapolations of the
considered observables can then be considered as an additional, systematic uncertainty of the obtained values.
The respective fit curves are displayed in \fig{fig:FiniteVolumeEffectsOfUpperHiggsMassBoundVEVandMH} and the 
corresponding infinite volume extrapolations of the renormalized vev and the Higgs boson mass, which 
have been obtained as the average over all presented fit results, are listed in 
\tab{tab:ResultOfUpperHiggsMassFiniteVolExtrapolation1}. 

The sought-after cutoff-dependent upper Higgs boson mass bound, and thus the main result of this paper, already presented in 
\fig{fig:UpperMassBoundFinalResult}, can then directly be obtained from the latter infinite volume extrapolation. 
The bounds arising in the full Higgs-Yukawa model and the pure $\Phi^4$-theory are jointly presented in 
\fig{fig:UpperMassBoundFinalResult}a. In both cases one clearly observes the expected decrease of the upper Higgs boson 
mass bound with rising cutoff $\Lambda$. Moreover, the obtained results can very well be fitted with the expected cutoff 
dependence given in \eq{eq:StrongCouplingLambdaScalingBeaviourMass}, as depicted by the dashed and solid curves in 
\fig{fig:UpperMassBoundFinalResult}a, where $A_m$, $B_m$ are the respective free fit parameters.

Concerning the effect of the fermion dynamics on the upper Higgs boson mass bound one finds in \fig{fig:UpperMassBoundFinalResult}a
that the individual results on the Higgs boson mass in the full Higgs-Yukawa model and the pure $\Phi^4$-theory at single cutoff 
values $\Lambda$ are not clearly distinguishable from each other with respect to the associated uncertainties. Respecting all 
presented data simultaneously by considering the aforementioned fit curves also does not lead to a much clearer picture, as can be 
observed in \fig{fig:UpperMassBoundFinalResult}a, where the uncertainties of the respective fit curves are indicated by the highlighted 
bands. At most, one can infer a mild indication from the presented results, being that the inclusion of the fermion dynamics causes 
a somewhat steeper descent of the upper Higgs boson mass bound with increasing cutoff $\Lambda$. A definite answer regarding the 
latter effect, however, remains missing here due to the size of the statistical uncertainties. The clarification of this issue would require 
the consideration of higher statistics as well as the evaluation of more lattice volumes to improve the reliability of 
the above infinite volume extrapolations.

On the basis of the latter fit results one can extrapolate the presented fit curves to very large values of the 
cutoff $\Lambda$ as illustrated in \fig{fig:UpperMassBoundFinalResult}b. It is intriguing to compare these large cutoff 
extrapolations to the results arising from the perturbative consideration of the Landau pole presented, for instance,
in \Ref{Hagiwara:2002fs}. One observes good agreement with that perturbatively obtained upper mass bound even though the 
data presented here have been calculated in the mass degenerate case and for $N_f=1$. This, however, is not too surprising 
according to the observed relatively mild dependence of the upper mass bound on the fermion dynamics. 

For clarification it is remarked that a direct quantitative comparison between the aforementioned perturbative and numerical 
results has been avoided here due to the different underlying regularization schemes. With growing values of $\Lambda$, however, 
the cutoff dependence becomes less prominent, thus rendering such a direct comparison increasingly reasonable in that limit. 

\includeFigTriple{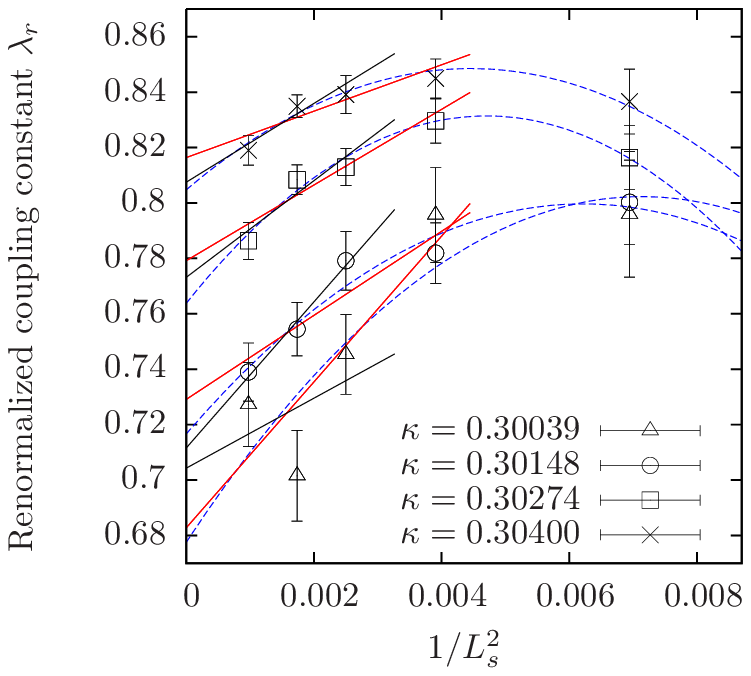}{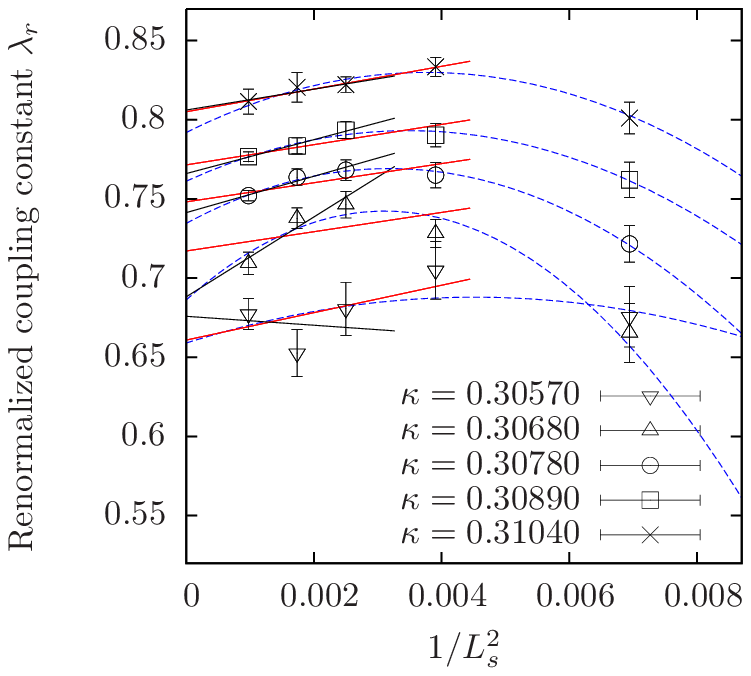}{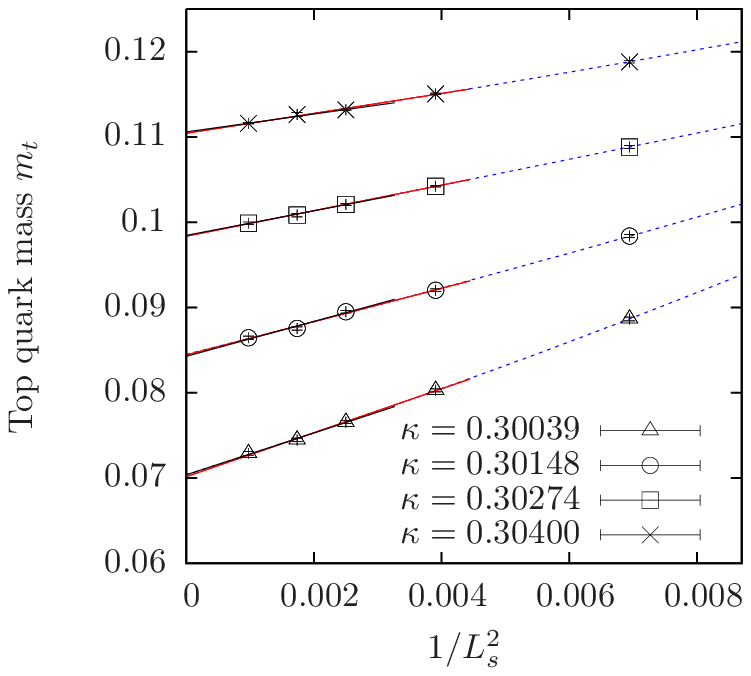}
{fig:FiniteVolumeEffectsOfUpperHiggsMassBoundLamR}
{The dependence of the renormalized quartic coupling constant $\lambda_r$ as well as the top quark mass $m_t$
on the squared inverse lattice side length $1/L^2_s$ 
is presented as calculated in the direct Monte-Carlo calculations specified in \tab{tab:SummaryOfParametersForUpperHiggsMassBoundRuns}.
Panels (a) and (c) show the results for the full Higgs-Yukawa model, while panel (b) refers to the pure $\Phi^4$-theory.
In all plots the dashed curves display the parabolic fits according to the fit ansatz in 
\eq{eq:ParaFit}, while the solid lines depict the linear fits resulting from \eq{eq:LinFit} for the two lower threshold 
values $L_s'=16$ (red) and $L_s'=20$ (black).
}
{Dependence of the renormalized quartic coupling constant $\lambda_r$ and the top quark mass $m_t$
on the squared inverse lattice side length $1/L^2_s$ at infinite bare quartic coupling constant.}

Furthermore, the question for the cutoff dependence of the renormalized quartic coupling constant $\lambda_r$ and -- in the case
of the full Higgs-Yukawa model -- the top quark mass with its associated value of the renormalized Yukawa coupling constant $y_{t,r}$
shall be addressed. For that purpose we follow exactly the same steps as above. The underlying finite volume lattice results on 
the renormalized quartic coupling constant and the top quark mass are fitted again with the parabolic and the linear fit 
approaches in \eq{eq:LinFit} and \eq{eq:ParaFit} as presented in \fig{fig:FiniteVolumeEffectsOfUpperHiggsMassBoundLamR}.

\includeTabHERE{|c|c|c|c|c|}{
\multicolumn{5}{|c|}{Renormalized quartic coupling constant $\lambda_r$} \\ \hline
$\kappa$  	 & $A^{(l)}_\lambda$, $L'_s=16$  & $A^{(l)}_\lambda$, $L'_s=20$         & $A^{(p)}_\lambda$         & $\lambda_r$               \\ \hline
$\,0.30039\,$  	 & $\,0.6827(280)\, $   & $\,0.7043(460)\, $  & $\,0.6775(452)\, $  & $\, 0.6882(406)(134)$  \\ 
$\,0.30148\,$  	 & $\,0.7291(118)\, $   & $\,0.7116(66) \, $  & $\,0.7166(134)\, $  & $\, 0.7191(110)(88)$  \\ 
$\,0.30274\,$  	 & $\,0.7791(79) \, $   & $\,0.7731(139)\, $  & $\,0.7638(81) \, $  & $\, 0.7720(103)(77)$  \\ 
$\,0.30400\,$  	 & $\,0.8164(71) \, $   & $\,0.8074(97) \, $  & $\,0.8047(67) \, $  & $\, 0.8095(79)(59)$  \\ 
$\,0.30570\,$  	 & $\,0.6609(182)\, $   & $\,0.6760(288)\, $  & $\,0.6590(288)\, $  & $\, 0.6653(258)(85)$  \\ 
$\,0.30680\,$  	 & $\,0.7171(201)\, $   & $\,0.6882(149)\, $  & $\,0.6862(182)\, $  & $\, 0.6972(179)(155)$  \\ 
$\,0.30780\,$  	 & $\,0.7482(56) \, $   & $\,0.7414(37) \, $  & $\,0.7346(24) \, $  & $\, 0.7414(41)(68)$  \\ 
$\,0.30890\,$  	 & $\,0.7716(47) \, $   & $\,0.7660(17) \, $  & $\,0.7612(34) \, $  & $\, 0.7663(35)(52)$  \\ 
$\,0.31040\,$  	 & $\,0.8051(23) \, $   & $\,0.8061(45) \, $  & $\,0.7919(88) \, $  & $\, 0.8010(59)(71)$  \\  \hline
\multicolumn{5}{|c|}{Top quark mass $m_{t}$} \\ \hline
$\kappa$  	 & $A^{(l)}_t$, $L'_s=16$  & $A^{(l)}_t$, $L'_s=20$         & $A^{(p)}_t$         & $m_{t}$               \\ \hline
$\,0.30039\,$  	 & $\, 0.0701(2) \, $   & $\, 0.0704(4) \, $  & $\, 0.0704(3) \, $  & $\, 0.0703(3)(2)$  \\ 
$\,0.30148\,$  	 & $\, 0.0844(3) \, $   & $\, 0.0843(6) \, $  & $\, 0.0845(4) \, $  & $\, 0.0844(5)(1)$  \\ 
$\,0.30274\,$  	 & $\, 0.0983(1) \, $   & $\, 0.0984(2) \, $  & $\, 0.0984(1) \, $  & $\, 0.0984(1)(1)$  \\ 
$\,0.30400\,$  	 & $\, 0.1104(1) \, $   & $\, 0.1106(1) \, $  & $\, 0.1105(2) \, $  & $\, 0.1105(1)(1)$  \\ 
}
{tab:ResultOfUpperHiggsMassFiniteVolExtrapolation2}
{The results of the infinite volume extrapolations of the Monte-Carlo data of the renormalized quartic coupling
constant $\lambda_r$ and the top quark mass $m_{t}$ are presented as obtained from the parabolic ansatz in 
\eq{eq:ParaFit} and the linear approach in \eq{eq:LinFit} for the considered lower threshold values $L_s'=16$ and $L_s'=20$. 
The final results on $\lambda_r$ and $m_{t}$, displayed in the very right column, are determined here by averaging over
the parabolic and the two linear fit approaches. An additional, systematic uncertainty of these final results is specified
in the second pair of brackets taken from the largest observed deviation among all respective fit results.}
{Infinite volume extrapolation of the Monte-Carlo data for the renormalized quartic coupling
constant $\lambda_r$ and the top quark mass $m_{t}$ at infinite quartic coupling constant.}

The corresponding infinite volume extrapolations are listed in \tab{tab:ResultOfUpperHiggsMassFiniteVolExtrapolation2},
where the final extrapolation result is obtained by averaging over all performed fit approaches. An additional systematic
error is again estimated from the deviations between the various fit procedures. 

The sought-after cutoff dependence of the aforementioned renormalized coupling constants can then directly be obtained
from the latter infinite volume extrapolations. The respective results are presented in \fig{fig:FinalResultsOnRenCoupAtStrongCoup} and
within the achieved accuracy one observes the renormalized coupling parameters to be consistent with the expected decline when increasing 
the cutoff $\Lambda$ as expected in a trivial theory. 
Again, the obtained numerical results are fitted with the analytically expected scaling behaviour given in 
\eq{eq:StrongCouplingLambdaScalingBeaviourLamCoupling} and \eq{eq:StrongCouplingLambdaScalingBeaviourYCoupling}. As already 
discussed for the case of the Higgs boson mass determination, the individual measurements of $\lambda_r$ in the two considered 
models at single cutoff values $\Lambda$ are not clearly distinguishable. Respecting the available data simultaneously by means of 
the aforementioned fit procedures also leads at most to the mild indication that the inclusion of the fermion dynamics results in a 
somewhat steeper descent of the renormalized quartic coupling constant with rising cutoff $\Lambda$ as compared to the pure $\Phi^4$-theory. 
A definite conclusion in this matter, however, cannot be drawn at this point due to the statistical uncertainties 
encountered in \fig{fig:FinalResultsOnRenCoupAtStrongCoup}.

Finally, the renormalized Yukawa coupling constant is compared to its bare counterpart depicted by the horizontal
line in \fig{fig:FinalResultsOnRenCoupAtStrongCoup}b. Since the latter bare quantity was chosen according to the tree-level
relation in \eq{eq:treeLevelTopMass} aiming at the reproduction of the physical top quark mass, one can directly infer from this
presentation how much the actually measured top quark mass differs from its targeted value of $\GEV{175}$. Here, one observes
a significant discrepancy of up to $\proz{2}$, which can in principle be fixed in follow-up lattice calculations, if
desired. According to the observed rather weak dependence of the upper Higgs boson mass bound on the Yukawa coupling constants,
however, such an adjustment would not even be resolvable with the here achieved accuracy. 

\includeFigDouble{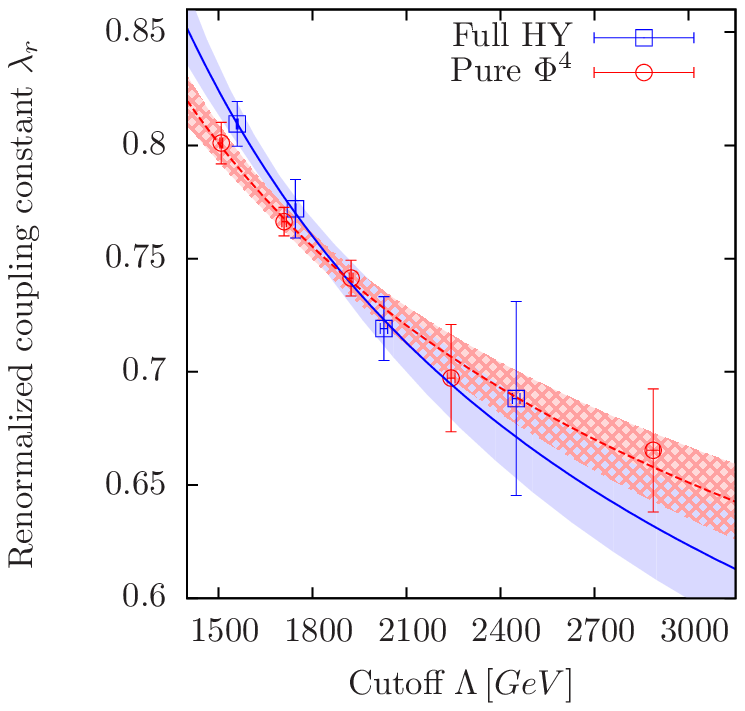}{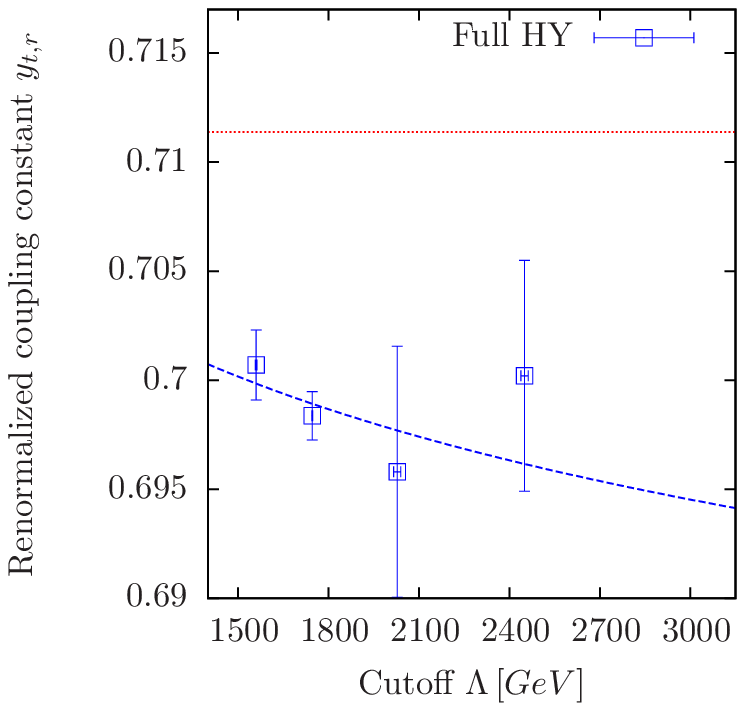}
{fig:FinalResultsOnRenCoupAtStrongCoup}
{The cutoff dependence of the renormalized quartic and Yukawa coupling constants is presented in panels (a) and (b), respectively,
as obtained from the infinite volume extrapolation results in \tab{tab:ResultOfUpperHiggsMassFiniteVolExtrapolation2}. The dashed 
and solid curves are fits with the respective analytically expected cutoff dependence in \eq{eq:StrongCouplingLambdaScalingBeaviourLamCoupling} 
and \eq{eq:StrongCouplingLambdaScalingBeaviourYCoupling}. The horizontal line in panel (b) indicates the bare degenerate Yukawa coupling 
constant underlying the performed lattice calculations. 
}
{Cutoff dependence of the renormalized quartic constant and the renormalized Yukawa coupling constant at infinite bare 
quartic coupling constant.}

\section{Summary and Conclusions}
\label{sec:conclusions}

The aim of the present work has been the non-perturbative determination of the cutoff-dependent upper mass bound of the 
Standard Model Higgs boson based on first principle computations, in particular not relying on additional information such as 
the triviality property of the Higgs-Yukawa sector.
The motivation for the consideration of the aforementioned mass bound finally lies in the ability of drawing conclusions
on the energy scale $\Lambda$ at which a new, so far unspecified theory of elementary particles definitely has to 
substitute the Standard Model, once the Higgs boson and its mass $m_H$ will have been discovered experimentally.
In that case the latter scale $\Lambda$ can be deduced by requiring consistency between the observed mass $m_H$
and the upper and lower mass bounds $\upBound$ and $\lowBound$, intrinsically arising from 
the Standard Model under the assumption of being valid up to the cutoff scale $\Lambda$. 

The Higgs boson might, however, very well not exist at all, especially since the Higgs sector can only be considered 
as an effective theory of some so far undiscovered, extended theory, due to its triviality property. In such a scenario, 
a conclusion about the validity of the Standard Model can nevertheless be drawn, since the non-observation of the Higgs 
boson at the LHC would eventually exclude its existence at energies below, lets say, $\TEV{1}$ thanks to the large accessible 
energy scales at the LHC. An even heavier Higgs boson is, however, definitely excluded without the Standard Model becoming 
inconsistent with itself according to the results in \sect{chap:ResOnUpperBound} and the requirement that the cutoff $\Lambda$ be 
clearly larger than the mass spectrum described by that theory. In the case of non-observing the Higgs boson at the LHC after
having explored its whole energy range, one can thus conclude on the basis of the latter results, that new physics must 
set in already at the TeV-scale.

For the purpose of establishing the aforementioned cutoff-dependent mass bound, the lattice approach has been employed to allow for a 
non-perturbative investigation of a Higgs-Yukawa model serving as a reasonable simplification of the full Standard Model, 
containing only those fields and interactions which are most essential for the Higgs boson mass determination. 
This model has been constructed on the basis of L\"uscher's proposals in \Ref{Luscher:1998pq} for the construction 
of chirally invariant lattice Higgs-Yukawa models adapted, however, to the situation of the actual Standard Model Higgs-fermion 
coupling structure, \ie for $\varphi$ being a complex doublet equivalent to one Higgs and three Goldstone modes. The resulting 
chirally invariant lattice Higgs-Yukawa model, constructed here on the basis of the Neuberger overlap operator, then obeys a global 
$\SUtwoTimesUoneY$ symmetry, as desired. 

The fundamental strategy underlying the determination of the cutoff-dependent upper Higgs boson mass bounds
has then been the numerical evaluation of the maximal Higgs boson mass attainable within the considered Higgs-Yukawa 
model in consistency with phenomenology. The latter condition refers here to the requirement of reproducing the phenomenologically
known values of the top and bottom quark masses as well as the renormalized vacuum expectation value $v_r$ of the scalar field, where 
the latter condition was used here to fix the physical scale of the performed lattice calculations. Owing to the
potential existence of a fluctuating complex phase in the non-degenerate case, the top and bottom quark masses
have, however, been assumed to be degenerate in this work. Applying this strategy requires the evaluation of the 
model to be performed in the broken phase, but close to a second order phase transition to a symmetric phase, 
in order to allow for the adjustment of arbitrarily large cutoff scales, at least from a conceptual point of view. 

As a first step it has explicitly been confirmed by direct lattice calculations that the largest attainable Higgs boson masses are 
indeed observed in the case of an infinite bare quartic coupling constant, as suggested by perturbation theory. Consequently, 
the search for the upper Higgs boson mass bound has subsequently been constrained to the bare parameter setting $\lambda=\infty$. 
The resulting finite volume lattice data on the Higgs boson mass turned out to be sufficiently precise to allow for their reliable 
infinite volume extrapolation, yielding then a cutoff-dependent upper bound of approximately $\upBound=\GEV{630}$ at a cutoff of 
$\Lambda=\GEV{1500}$. These results were moreover precise enough to actually resolve their cutoff dependence as demonstrated in 
\fig{fig:UpperMassBoundFinalResult}, which is in very good agreement with the analytically expected logarithmic decline, and thus 
with the triviality picture of the Higgs-Yukawa sector. 

It is remarked here, that this achievement has been numerically demanding, since the latter logarithmic decline of the upper bound 
$\upBound$ is actually only induced by subleading logarithmic contributions to the scaling behaviour of the considered model close to 
its phase transition, which had to be resolved with sufficient accuracy. By virtue of the analytically expected 
functional form of the cutoff-dependent upper mass bound, which was used to fit the obtained numerical data, an extrapolation 
of the latter results to much higher energy scales could also be established, being in good agreement with the corresponding 
perturbatively obtained bounds~\cite{Hagiwara:2002fs}. A direct comparison has, however, been avoided due to the different
underlying regularization schemes.

The interesting question for the fermionic contribution to the observed upper Higgs boson mass bound has then been addressed
by explicitly comparing the latter findings to the corresponding results arising in the pure $\Phi^4$-theory. For the considered
energy scales this potential effect, however, turned out to be not very well resolvable with the available accuracy of the lattice 
data. The performed fits with the expected analytical form of the cutoff dependence only mildly indicate the upper mass bound in the 
full Higgs-Yukawa model to decline somewhat steeper with growing cutoffs than the corresponding results in the pure $\Phi^4$-theory. 
To obtain a clearer picture in this respect, higher accuracy of the numerical data and thus higher statistics of the underlying 
field configurations would be needed.

\section*{Acknowledgments}
We thank J. Kallarackal for discussions and M. M\"uller-Preussker for his continuous support.
We are grateful to the "Deutsche Telekom Stiftung" for supporting this study by providing a Ph.D. scholarship for
P.G. We further acknowledge the support of the DFG through the DFG-project {\it Mu932/4-1}.
The numerical computations have been performed on the {\it HP XC4000 System}
at the {\it Scientific Supercomputing Center Karlsruhe} and on the
{\it SGI system HLRN-II} at the {\it HLRN Supercomputing Service Berlin-Hannover}.

\bibliographystyle{unsrtOWN}
\bibliography{UpperHiggsBosonMassBounds}

\end{document}